\documentclass{aastex62}
\usepackage{textcomp}
\usepackage{graphicx}
\usepackage{here}
\usepackage{hhline}
\usepackage{longtable}

\def\hoge<#1>{\langle #1 \rangle}

\received{July 4, 2018}
\revised{October 1, 2018}
\accepted{October 22, 2018}
\submitjournal{AJ}

\shortauthors{Goda \& Matsuo}

\begin{document}

\title{A New Method for Calibration of Gain Variation in Detector System}

\author{Shohei Goda}
\affil{\rm Department of Earth and Space Science, Graduate School of Science, Osaka University, 1-1, Machikaneyamacho, Toyonaka, Osaka 560-0043, Japan}

\author{Taro Matsuo}
\affiliation{\rm Department of Earth and Space Science, Graduate School of Science, Osaka University, 1-1, Machikaneyamacho, Toyonaka, Osaka 560-0043, Japan}

\begin{abstract}

Transit spectroscopy of habitable planets orbiting late-type stars requires high relative spectro-photometric accuracy between wavelengths during transit/eclipse observation. The spectro-photometric signal is not affected only by image movement and deformation due to wavefront error but also by electrical variation in the detector system. These time-variation components, coupled to the transit signal, distort the measurements of atmospheric composition in transit spectroscopy. Here we propose a new concept for improvement of spectro-photometric accuracy through the calibration of the time-variation components in the detector system by developing densified pupil spectroscopy that provides multiple spectra of the star-planet system. Owing to a group of pixels exposed by the object light (i.e., science pixels), pixel-to-pixel variations can be smoothed out through an averaging operation, thus only common time-variation components over the science pixels remain. In addition, considering that the detector plane is optically conjugated to the pupil plane, a pupil mask can completely block astronomical light incoming into residual pixels. The common time-variation components are reconstructed with the residual pixels and reduced into a random term. Applying the densified pupil spectrograph with a mid-infrared detector system to a large space cryogenic telescope such as the Origins Space Telescope, we show that the system nearly achieves photon-noise-limited performance and detects absorption features through transmission spectroscopy and secondary eclipse of terrestrial planets orbiting M-type stars at 10 pc with 60 transit observations. Thus, the proposed method contributes to the measurement of planetary habitability and biosignatures of the nearby transiting habitable candidates.

\end{abstract}

\vspace{1cm}

\keywords{instrumentation: spectrographs -- methods: data analysis -- techniques: spectroscopic -- planets and satellites: atmospheres -- planets and satellites: terrestrial planets}

\section{Introduction} \label{sec:intro}

Characterization of Earth-like planets' atmospheres in the mid-infrared wavelength is an important approach toward search for life in the universe. A sign of a non-equilibrium atmosphere that indicates biological activity on a planet can be confirmed through measurements of both oxidized species such as $\rm O_2$ and $\rm O_3$ and reduced species such as $\rm CH_4$ and $\rm N_{2}O$ \citep{1965Natur.207..568L, 1993Natur.365..715S}. Most of these species produce strong absorption lines in the mid-infrared wavelength range: $\rm O_3$ at 9.6 \textmu m, $\rm CH_4$ at 3.3 and 7.7 \textmu m, and $\rm N_{2}O$ at 7.8 \textmu m. Note that the thermodynamic disequilibrium atmosphere with gases of $\rm CO$ and $\rm CH_4$ can be formed through abiotic process at room temperature \citep{2014PNAS..11112641K}; the combination of $\rm O_2$ ($\rm O_3$) and $\rm CH_4$ or $\rm O_2$ ($\rm O_3$) and $\rm N_{2}O$ is a reliable indicator of biological activity \citep[e.g.,][]{2017ARA&A..55..433K}. The ozone gas is also a good tracer for the presence of life on a planet whose surface is undergoing an oxidization process because the deep absorption at 9.6 \textmu m is formed even under low oxygen concentration in the atmosphere \citep{2018ApJ...854...19R}. Considering the oxidation history of the Earth surface by oxygenic photosynthesis \citep[e.g.,][]{2007Natur.448..1033, 2014Natur.506..307L}, we can investigate whether the oxidation history on a planet’s surface corresponds to biological activity similar to that of Earth.

There are mainly two approaches for atmospheric spectroscopy of habitable terrestrial planets: direct imaging that directly detects the planet’s light embedded in the bright halo of a star \citep[e.g.,][]{2007Natur.446..771T, 2006Natur.442...51C}, and transit spectroscopy that detects additional absorptions by atmospheric molecules of a transiting planet while the planet passes in front of the primary star \citep[e.g.,][]{2002ApJ...568..377C, 2008Natur.452..329S}. In addition, we can detect the emissions by making the difference between before/after and during the eclipse \citep[e.g.,][]{2005Natur.434..740D, 2008ApJ...686.1341C}. Direct imaging in the mid-infrared wavelength requires a nulling interferometer that comprises two collecting apertures with a baseline of a few hundred meters \citep[e.g.,][]{1978Natur.274..780B, 1997ApJ...475..373A, 2011ApJ...729...50M}. This separation improves the inner working angle as well as the spatial resolution in that wavelength range by introducing an achromatic phase shift of π between the two beams so that the stellar light is suppressed. On the other hand, although the transit spectroscopy can be performed only when an observing condition is met, in which an observer, a host star, and its orbiting planet are aligned, we do not essentially need an interferometer for higher spatial resolution because the transit observation is independent of the angular distance between the host star and its planet. The species related to the biomarkers and planetary habitability in the stratosphere of the Earth also produce absorption lines in mid-infrared range \citep{2009ApJ...698..519K}; these gases increasingly tend to be detected from the transmission spectrum \citep[e.g.,][]{2016MNRAS.461L..92B, 2011Cambridge.v1}. Furthermore, since the contrast between the cool sunspot or bright facula and the stellar photosphere become smaller as the wavelength is longer, the systematic error arising from the stellar activity can be reduced in that range \citep{2012A&A...539A.140B}. Thus, the transit spectroscopy in the mid-infrared wavelength opens a new path toward characterization of the Earth-like planets.

Transit observations in the mid-infrared region have been conducted by the Spitzer space telescope and have revealed the temperature distribution and atmospheric circularization on the surface of the transiting planets \citep[e.g.,][]{2007Natur.447..183K, 2014Sci...346..838S, 2016ApJ...821....9K} as well as the atmospheric composition of hot Jupiters \citep[e.g.,][]{2007Natur.448..169T, 2009ApJ...707...24M}. The James Webb Space Telescope (JWST), as the successor to the Spitzer space telescope, is able to perform transit spectroscopy of sub-Neptunes and super-Earths owing to its large aperture \citep[e.g.,][]{2009PASP..121..952D, 2016ApJ...817...17G}. However, since the spectro-photometric accuracy of JWST in the mid-infrared region is expected to be limited by several dozen ppm due to various instrumental systematic effects even under small telescope pointing jitter \citep{2014PASP..126.1134B}, it is difficult (but not impossible) to implement spectroscopy of nearby terrestrial planets in the habitable zones around late-type stars such as the Trappist-1 system \citep{2017Natur.542..456G} through co-addition of the multiple transmission spectra \citep{2018arXiv180303730B}. Although several advanced data analysis for the space telescopes, including the Spitzer space telescope, has been proposed \citep[e.g.,][]{2013ApJ...766....7W, 2016ApJ...820...86M}, whether such systematic errors can be reduced down to 10 ppm with the sophisticating techniques is still unknown. Thus, the instrumental systematic components should be carefully treated and accurately calibrated for transit spectroscopy of the Earth-like planets with future space telescopes.

Image movement and deformation of the point-spread function (PSF) on intra- and inter-pixel sensitivity variations \citep{2007PASP..119..466B, 2012SPIE.8442E..1YI} due to the telescope pointing jitter and wavefront errors generates time-variable components that affect the observational data from space \citep{2014ASPC..485..407C, 2014ApJ...790...53Z}. This occurs because the spectrum is formed on the focal plane for general-purpose spectrographs and the wavefront errors on the pupil plane directly deform the point-spread function. Densified pupil spectroscopy has been proposed for minimization of the instrumental systematic noise caused by wavefront errors through acquiring multiple spectra of the primary mirror on the detector plane \citep{2016ApJ...823..139M}. Pupil masks that form apodized PSFs on a field-stop have been also proposed to reduce the photometric variation arising from the field-stop loss coupled with the pointing jitter and wavefront errors \citep{2017AJ....154...97I}.

The systematic errors are generated by electric instability in a semiconductor devise and a Read-Out Integrated Circuit (ROIC), as well as the PSF movement and deformation due to the wavefront errors in the optical system. According to the previous studies on the semi-conductor devices and ROICs developed for the mid-infrared observations, the systematic component related to the electric instability is mainly divided into three. The first photometric variation is observed as a steep ramp of a small percentage at the beginning of the transit observations and a fall of 0.1 percentage point over a timescale of 10-20 hours after the ramp \citep[e.g.,][]{2003SPIE.4850...98Y, 2012ApJ...752...81C}. We hereafter name the long-term fall of the detector gain as “fallback.” Note that this phenomenon does not always occur but we consider the signal with this effect in the present paper. The ramp is thought to be caused by the variation of the ratio of the number of incident photons to that of photoelectrons that arrives to the ROIC, i.e., the effective gain variation \citep{2009ApJ...703.1884S, 2009ApJ...703..769K, 2010ApJ...721.1861A}. The effective gain variation arises from the change of the number of photoelectrons trapped due to detector impurities; as the photoelectrons bury the holes produced by the detector impurities, they increasingly tend to arrive to the ROIC without being captured by the holes. This phenomenon is called “electron trapping.” The time-variation component due to the electron trapping can be reduced down to several hundred ppm by pre-flashing, which irradiates the detector plane with bright stellar light before transit observations, such that the holes are buried by the photoelectrons generated due to the bright object \citep{2009ApJ...703.1884S, 2009ApJ...703..769K}. Regarding the fallback after the steep ramp, although what is behind the fallback is still unknown, the fallback may be related to the persistent effect in the multiplexers of the ROIC \citep{2012ApJ...752...81C}. Note that there are some cases where no ramp and fallback is observed in the transit phase curve \citep[e.g.,][]{2007Natur.447..183K}. The latent effect also leads to a potential systematic error. When a bright source is observed, the bright source leaves its footprint on the detector even after it is removed. A latent level of about 1 \% for the bright source was reported for the detector systems developed for the mid-infrared instrument (MIRI) of the JWST \citep{2015PASP..127..675R}. Although the cause of the latent image is not fully understood, the latent effect is thought to be generated by capturing photoelectrons in the blocking layer of the semiconductor device and dielectric relaxation because of its long RC time constant \citep{2015PASP..127..665R}. The third systematic error arises from fluctuations of the bias voltage acting on the semiconductor and the constant voltages acting on the multiplexers of the ROIC. This occurs because the detector responsivity strongly depends on the bias voltage \citep{2015PASP..127..665R} and the fluctuation of the constant voltages leads to the gain variation in the ROIC. Since a source follower is installed in each pixel unit of the complementary MOS (CMOS) infrared detector system, there are individual noises that depend upon the pixels. Additionally, because the bias voltages are applied to every four pixels for the Si:As detectors developed for JWST/MIRI, gain variations are induced in the pixels and lead to pixel-to-pixel time-variation if the bias voltages themselves vary in time. On the other hand, when time-variation of the bias voltage exceeds the full-frame readout time, the temporal change in the bias voltage commonly affects the signals of all pixels. Therefore, high-frequency fluctuation of the bias voltage causes the gain variation between the pixels, and the common time-variation of the gain over the entire detector occurs due to the low-frequency fluctuation. Note that, although the fluctuation in the low-frequency is observed in the output signal as 1/f noise, it is difficult to perfectly separate the detector noise from the astronomical signal \citep{2014JWST.003852..12, 2015PASP..127.1144R}.

Based on the above background, we propose a new method for correction of the time-variation present in the semiconductor devices and ROIC, developing densified pupil spectroscopy. First, the light from an object is dispersed into a number of spectra by the densified pupil spectrograph and a large number of pixels are exposed by the object light. The exposed pixels are named “science pixels.” The pixel-to-pixel time-variation components are smoothed out through average of the science pixels and the common components over the entire detector are extracted. Second, because the number of photons falling on a pixel is largely reduced, the steep ramp at the beginning of the transit observation can be mitigated if the ramp is caused due to electron trapping \citep[e.g.,][]{2003SPIE.4850...98Y, 2012ApJ...752...81C}. The latent effect can be also reduced owing to a small number of the photoelectrons. Third, we place a cold photon-shield mask on which the densified sub-pupils are formed to block any thermal light. A number of the cold pixels, in which only the dark current generates, are produced. We refer the cold pixels as “reference pixels.” Since the value of the dark current is accurately proportional to the temperature of the detector \citep[e.g.,][]{2003SPIE.4850..890E}, the fluctuations of the reference pixels reflect the time-variation generated in the detector system if the detector temperature is stable over the transit observation. Owing to the extraction of the common time-variation components over the entire detector by each average of the science and reference pixels, the long-term fluctuations acting on the science pixels can be calibrated by simultaneous measurement of the object and the reference signals. Note that several methods for calibration of the gain fluctuation with reference pixels, which are not sensitive to an object light and are different from those introduced in this paper, have been proposed \citep[e.g.,][]{2007ARA&A..45...77R, 2017PASP..129j5003R}.

In this paper, we propose a concept that mitigates and calibrates the systematic noises present in the mid-infrared wavelength range of detector systems used to characterize nearby transiting terrestrial planets. We briefly introduce the method for calibration of the systematic components and an optical design that enables the application of the calibration technique to spectro-photometric data in Section \ref{sec:concept}. Next, we describe the detailed analytical expression for the mid-infrared detector systems in Section \ref{sec:analysis}. Assuming that the densified pupil spectrograph with the mid-infrared detector system is applied to the Origins Space Telescope \citep{2018arXiv180307730F}, we perform numerical simulations and then show the results in Section \ref{sec:simulation}. Finally, we raise several potential problems that are not considered in the numerical simulations and discuss the impacts on transit spectroscopy in Section \ref{sec:discussion}.

\section{Concept} \label{sec:concept}

In this section, we introduce a conceptual design for the calibration of the time-variation components present in a detector system composed of a semiconductor device and a readout integration circuit (ROIC), developing the densified pupil spectrograph. We also express the time-variation components with simple equations and show how the observational data acquired through the new design is calibrated and approaches a true astronomical value that is free from the systematic components.

\subsection{Design} \label{subsec:design}

When the numbers of the photons from an astronomical source and a background component, $f_s$ and $f_z$, respectively, fall on an (i, j) pixel in a detector system with a quantum efficiency of $\eta$ and a dark current corresponding to the number of electrons, $Q$, the observational data $D$ for the (i, j) pixel is written as a function of time, $t$:
\begin{equation} \label{equ:raw}
D_{ij}(t) = A_{ij}(t)\beta(t,f_{ij})[\eta_{ij}\{\alpha^{'}_{ij}(WFE(t))f_{s,ij}(t)+\alpha_{ij}f_{z,ij}\}+Q_{ij}]+V_{ij} ,
\end{equation}
where $A$ is an effective gain that yields a conversion factor from electron to voltage (gain) in the detector system, $\beta$ is another effective gain that indicates the fraction of the photoelectrons arriving at the ROIC to those generated in the semiconductor device, and $\alpha^{'}$ and $\alpha$ are the effective gains for a point source and a diffuse background, respectively. While the former effective pixel gains are determined by both the wavefront error ($WFE$) in the optical system and intra- and inter-pixel sensitivity variation \citep[e.g.,][]{2007PASP..119..466B, 2014ASPC..485..407C}, the latter is characterized by only the inter-pixel map. $V$ is an offset voltage present in the readout electronics. Most of the factors shown in Equation (\ref{equ:raw}) include time-variation as well as random components. The time-variation of the effective gain, $A$, is caused by fluctuations of the bias voltage supplied to the semiconductor device and the constant voltages supplied to the ROIC. For detectors without a global shutter such as the MIRI detectors developed for the JWST, while the high-frequency fluctuation produces pixel-to-pixel time-variations, the low-frequency one leads to a common time-variation over the entire detector. This is because the high-frequency fluctuation of the bias voltage leads to a different voltage being applied to each pixel while the low-frequency fluctuation provides a constant voltage over the entire detector. Another effective gain, $\beta$, introduces a ramp at the beginning of the transit observation and a fallback over a timescale of 10-20 hours after the ramp, which are thought to be caused by the trapping of photoelectrons at holes due to detector impurities and the persistence effect of the ROIC. The effective pixel gains for a point source, $\alpha^{'}$, also contributes to the photometric variation because the wavefront error in the telescope and optical system leads to the image movement and deformation on the intra- and inter-pixel variation. In the mid-infrared wavelength region, the zodiacal light contributes to the transit light curve as an offset component, as well as photon noise. Note that we assume that there is no variation of the background light from the zodiacal light during the transit observation. We also assume that $Q$ is a constant dark current and $V$ is reduced into a random readout noise by a sophisticated detector sampling such as correlated double sampling (CDS) and fowler sampling \citep{2016Wiley.v2}. Thus, the transit light curve related to $f(t)$ is mainly affected by the time-variations of the two effective gains, $A(t)$ and $\beta (t)$, the effective pixel gain, $\alpha$, and the background component $f_z$, for general-purpose spectrographs.

\begin{figure}[H]
\begin{center}
\includegraphics[angle=270, width=12cm]{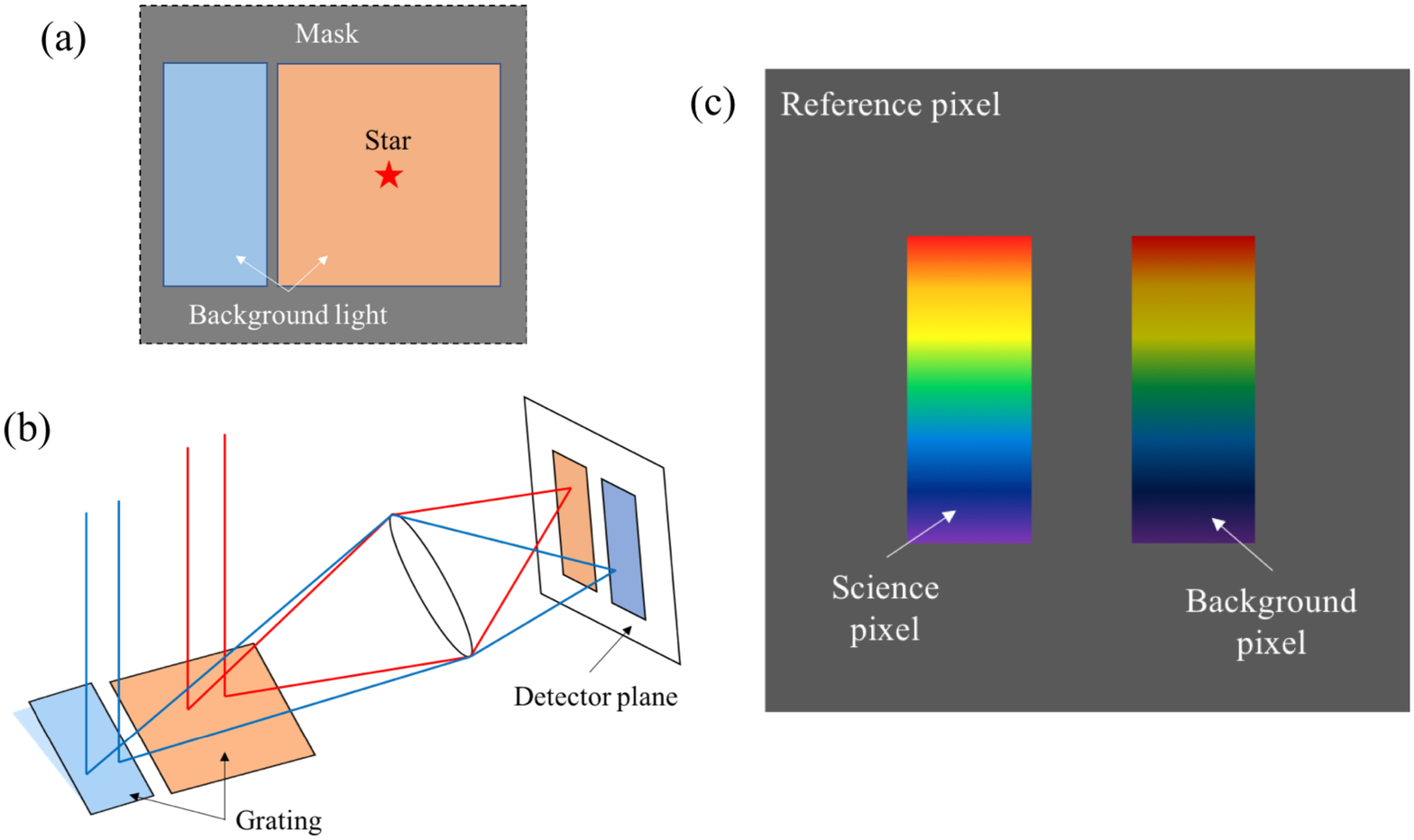}
\caption{(a) Orange and blue regions indicate the two field-of-views corresponding to the science pixels and background pixels, respectively. (b) Conceptual design of two diffraction gratings placed on the focal plane. (c) Example of pixel mapping on the detector plane for the modified design of the densified pupil spectrograph. There are three types of pixels: science pixels (bright rainbow), background pixels (dark rainbow) and reference pixels (black). The horizontal and vertical directions correspond to the spatial position of the pupil and the wavelength dispersion, respectively. The numbers of the science and background pixels equal and are determined by the number of pupil slicers and the densification factor of the pupil size. \label{fig:detector_plane}}
\end{center}
\end{figure}

Here, we propose a design for minimizing and calibrating the various time-variation components shown in Equation (\ref{equ:raw}) developing the densified pupil spectroscopy concept that forms multiple spectra on the detector plane optically conjugated to the primary mirror of a telescope \citep[see also Figure \ref{fig:detector_plane}]{2016ApJ...823..139M}. First, because the densified pupil spectrograph provides the spectra of the divided primary mirror, the wavefront error present in the telescope and optical system does not lead to the image movement and deformation of the detector plane; therefore, $\alpha^{'}(WFE)$ for a point source becomes equal to that for a diffuse background, $\alpha$. Second, owing to the multiple spectra formed on the number of the pixels, the pixel-to-pixel time-variation components with high frequency can be smoothed out through the average of the science pixels that are exposed by the object spectra. According to the transit spectrograph designed for the Origins Space Telescope \citep{2016SPIE.10698}, the number of the samplings for each spectral element with a spectral resolution of 100 is around 2000. Thus, the common long-term variation to the entire detector is extracted from the averaged spectrum. Last, since the number of photons falling on each science pixel decreases, the timescale of the electron trapping decrease is much longer  than that for a conventional spectrograph; hence, the long-term variation component due to the trapping effect can be approximately ignored. Note that we quantitatively evaluate the impact of the trapping effect on the transit spectroscopy in Section \ref{sec:discussion}. Based on the above considerations, Equation (\ref{equ:raw}), under the densified pupil spectrograph, is rewritten as
\begin{equation}
D_{ij}(t) \approx A_{ij}(t)[\alpha_{ij}\eta_{ij}\{f_{s}(t)+f_{z}\}+Q_{ij}]+\sigma_{read,ij} ,
\end{equation}
where $\alpha^{'}_{i,j}(WFE(t))$ becomes $\alpha_{i,j}$ for the densified pupil spectrograph, and $\sigma_{read,ij}$ is the readout noise generated through the reduction of the offset voltage with CDS or the fowler sampling. In addition, the numbers of the photons from the target object and background component, $f_{s,ij}(t)$ and $f_{z,ij}$, are replaced by $f_{s}(t)$ and $f_z$, respectively, that do not depend on the detector pixels exposed by the object light because the densified pupil spectrograph provides multiple equivalent rectangular spectra. The averaged signal over the science pixels is
\begin{equation}
\hoge<D_{ij}(t)>_{pix} \approx A_{com}(t)[\hoge<\alpha_{ij}\eta_{ij}>_{pix}\{f_{s}(t)+f_{z}\}+\hoge<Q_{ij}>_{pix}]+\sqrt{\hoge<\sigma^{2}_{read,ij}>_{pix}} ,
\end{equation}
where $\hoge<X>_y$ indicates the averaged $X$ over $y$ and $A_{com}$ represents the common time-variation gain over the entire detector, $\hoge<A_{ij}(t)>_{pix}=A_{com}(t)$. Thus, when the densified pupil spectrograph is applied to the transit observations, the common low-frequency time-variation components, $A_{com}$, impact both the transit-light curve and the background light, $f_z$, and the dark current, $Q$.

\begin{figure}[H]
\begin{center}
\includegraphics[width=12cm]{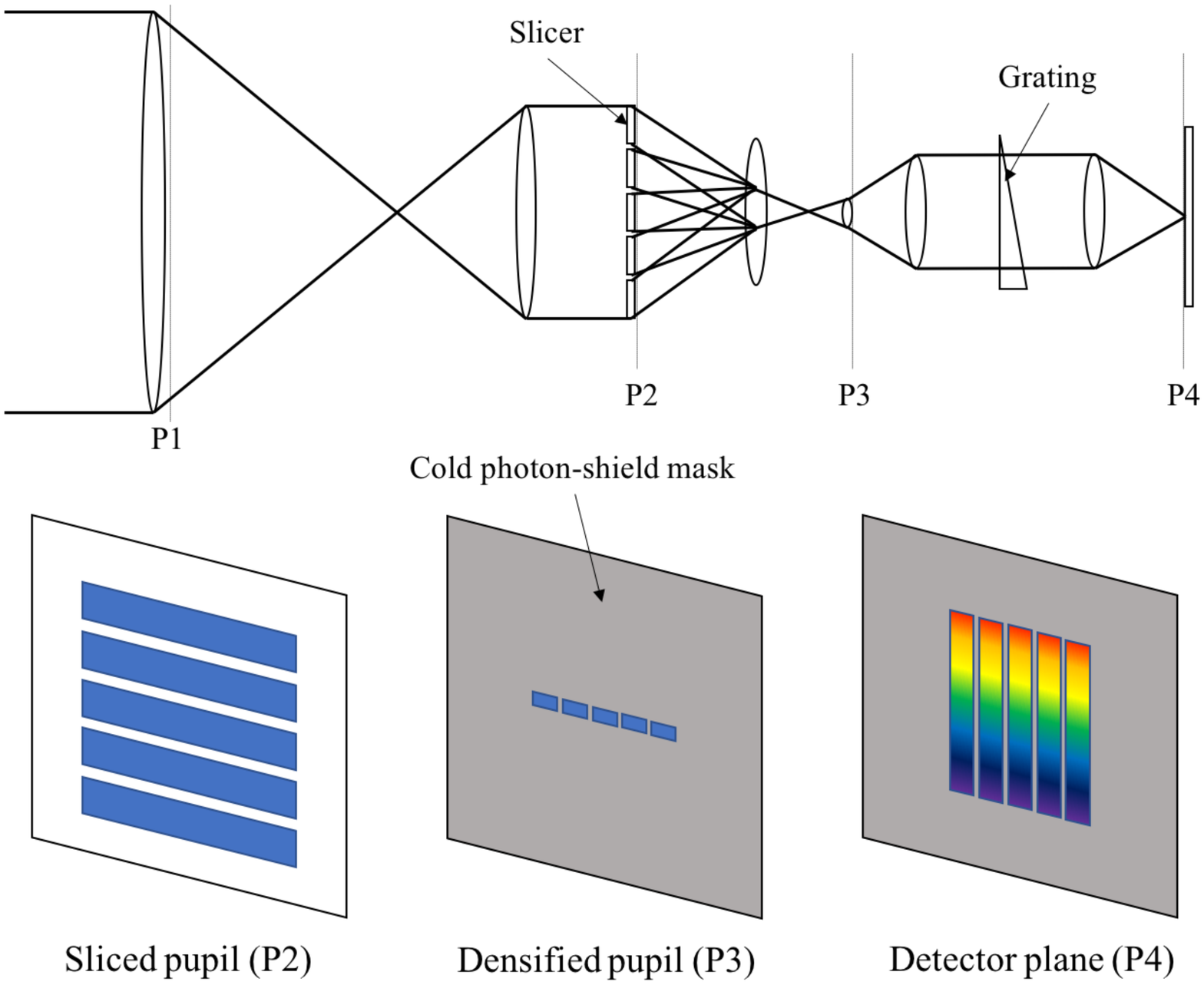}
\caption{Conceptual design of the densified pupil spectrograph (top). P1 plane is the primary mirror of a telescope. P2, P3, and P4 planes correspond to the pupil that is optically conjugated to the primary mirror P1. Incident light is first divided with a pupil slicer on P2 plane. After that, the divided sub-pupils are rearranged in line and each sub-pupil is densified on P3 plane. The densified sub-pupils are diffracted as a point-source and are then dispersed with a diffraction grating. The spectra of the sub-pupils are formed on the detector plane P4. \label{fig:spectrograph}}
\end{center}
\end{figure}

In addition to the pixels exposed by the target light (i.e., science pixels), two types of pixels are prepared for calibration of the time-variation components and subtraction of the offset component. One is referred to as “reference pixel”, which is generated by placing a photon-shield mask on a pupil plane (Figure \ref{fig:spectrograph}, P3) to block any thermal light and in which only dark current is generated. As discussed in the next section, because the dark current is also affected by the common time-variation components over the entire detector, the low-frequency systematic component can be reconstructed from the reference pixels. The other is called “background pixel.” The observational field does not include only the host star and planet but also the background light (Figure \ref{fig:detector_plane}, (a)). Because the zodiacal light is very bright over the entire sky in the mid-infrared wavelength \citep[e.g.,][]{1998ApJ...508...44K, 2016AJ....151...71K}, the zodiacal light contributes to the transit signal as an offset, as well as photon noise. In order to remove the background component, the two diffraction gratings on the focal plane are used and tilted differently so that the sum of the target and background light, and only the background light, are focused at two different positions on the detector plane (see Figure \ref{fig:detector_plane}, (b)). The background pixels are distributed oner the same area as the science ones (Figure \ref{fig:detector_plane}, (c)).

There is a noncommon path error between the science and background pixels, which may affect the subtraction of the offset component from the science pixels. However, the individual difference of the two gratings can be ignored because the large-diameter science light enters the two gratings owing to the densified pupil spectrograph, leading the local difference between the two gratings to be smoothed out. In addition, even though the second grating is tilted relative to the astronomical light, the characteristic of the densified pupil spectroscopy remains; the noise floor of the background light is the same as that of the science light. On the other hand, because the background light is formed at the different locations on the detector plane, the flat-field error affects the zodiacal light differently from how it influences the science light. One possible way to mitigate the impact of the flat-field error on the proposed method is to observe only the zodiacal light both with the science and background pixels and to calibrate the difference between them before the science observations. Even if the residual offsets still remain through the calibration, they can be eliminated by extracting the spectrum of the transiting planetary atmosphere (i.e., subtraction of the signal before and after the transit from that during the transit). This is because the residual offset is not expected to change over a single transit observation. Note that we quantitatively discuss the impact of the flat-field uncertainty upon the proposed method in Section \ref{sec:simulation}.

\subsection{Method} \label{subsec:method}

\begin{figure}[H]
\begin{center}
\includegraphics[width=12cm]{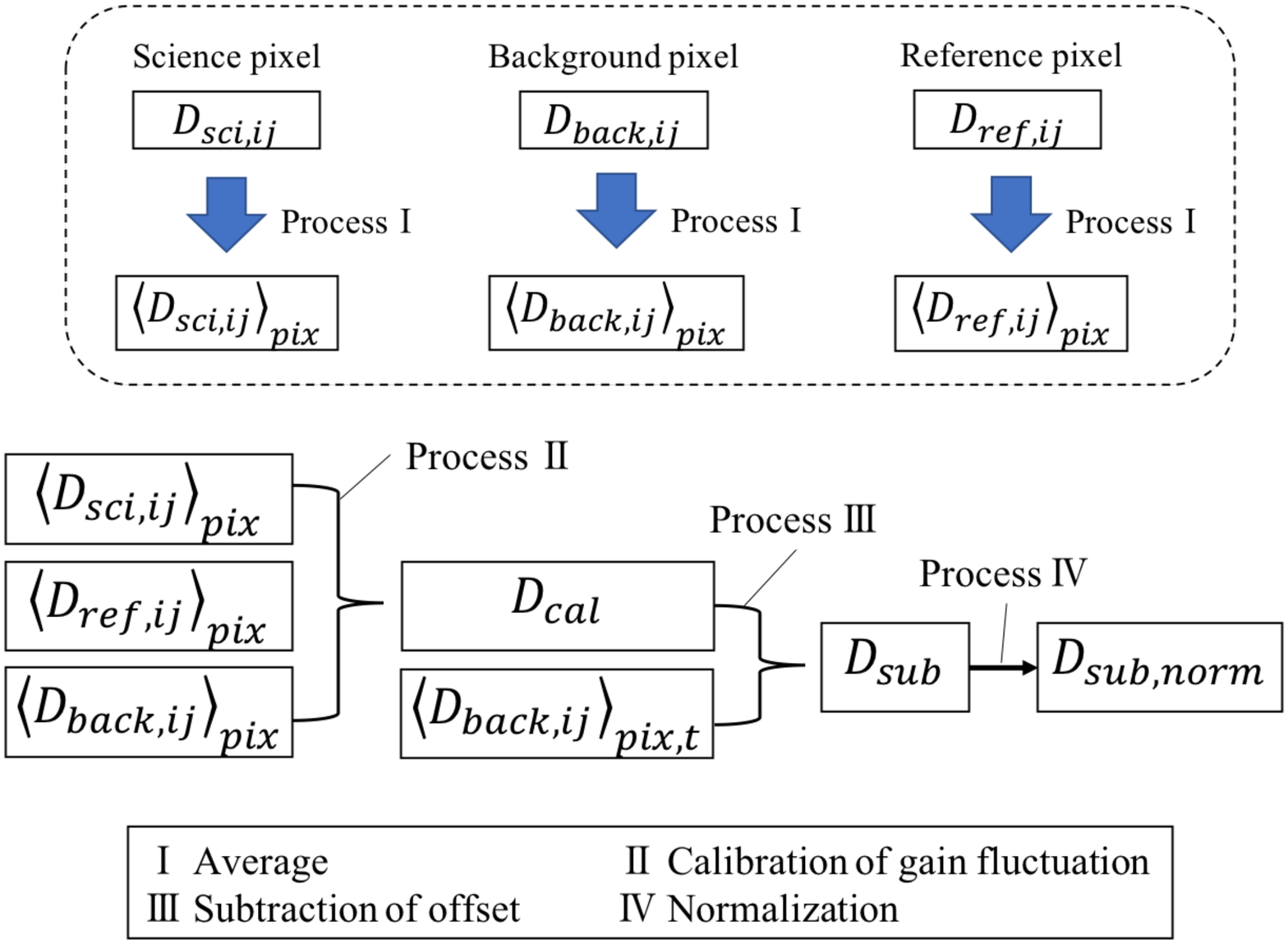}
\caption{Flowchart of the calibration method. \label{fig:calibration_flowchart}}
\end{center}
\end{figure}

We briefly introduce the procedure for the proposed method in this subsection. The reference and background pixels that are generated by the modified design of the densified pupil spectrograph are applied for calibration of the systematic components present in the semiconductor device and the ROIC. The signals of the science, background, and reference pixels, $D_{sci,ij}$, $D_{back,ij}$ and $D_{ref,ij}$, are, respectively, written as
\begin{equation}
D_{sci,ij}(t) \approx A_{ij}(t)[\alpha_{ij}\eta_{ij}\{f_{s}(t)+f_{z}\}+Q_{ij}]+\sigma_{read,ij} ,
\end{equation}
\begin{equation}
D_{back,ij}(t) \approx A_{ij}(t)(\alpha_{ij}\eta_{ij}f_{z}+Q_{ij})+\sigma_{read,ij} ,
\end{equation}
and
\begin{equation}
D_{ref,ij}(t) \approx A_{ij}(t)Q_{ij}+\sigma_{read,ij} .
\end{equation}
We average the signals over these pixels to smooth out the pixel-to-pixel variations (Process I). An overview of the method for calibration of gain variation in the detector system is shown in Figure \ref{fig:calibration_flowchart}. Subsequently, the calibration of the common time-variation component is implemented using the reference and background pixels. Since the time-variation component is common to the entire detector pixels, including the reference and background pixels, as well as the science pixels, it is possible to reconstruct the common systematic components from the reference and background data because the reference and background pixels are not affected by the transit signal but by the common systematic components. The systematic components associated to the reference and background data are derived by subtraction of the time- and pixel-averaged ones from the pixel-averaged ones at each frame. Given that there is linearity among the three types of pixels, the calibrated signal, $D_{cal}$, is acquired by subtraction of the common time-fluctuation components without a transit signal from the average value over the science pixel (Process I\hspace{-.1em}I). Last, using the time- and pixel-averaged background pixel value, the offset component can be also subtracted; $D_{sub}$ is acquired (Process I\hspace{-.1em}I\hspace{-.1em}I), and we can obtain the normalized transit signal, $D_{sub,norm}$, using the subtracted data (Process I\hspace{-.1em}V). Thus, we can obtain the calibrated signal that is free from both the systematic and offset components. Note that we assumed the dark current and background light to be constant during the transit observation. The proposed method requires stabilization of the detector's temperature and calibration of the dark current value using the time series of the detector temperature. On the other hand, we also discuss cases that the dark current and the zodiacal light have time-variation fluctuations, and evaluate how the time-variation components of the dark current and zodiacal light affect the proposed method in Section \ref{subsec:time_variation}. In Section \ref{sec:analysis}, we show a detailed mathematical description of the conceptual design optimized for CMOS Infrared detector systems.

\section{Mathematical analysis for infrared detector systems} \label{sec:analysis}

In this section, we apply the proposed method for calibration of the long-term variation components to infrared detector systems and then show a mathematical description of the method optimized for the infrared detector systems.
In this paper, we refer two papers on the detector sampling and readout integration circuit developed for the infrared detector systems \citep{2009ITED...56.2506J, 2015PASP..127..675R}.

The detector system operating in the infrared wavelength range is mainly divided into three subsystems: a semiconductor device in which photons enter and are converted to photoelectrons, a ROIC in which the photoelectrons are converted to voltage in each pixel unit and voltages of the entire detector pixels are combined into one or a few signal lines through one or more multiplexers, and several non-cryogenic electronics in which the combined voltages are transferred to digital values with analog to digital converters (ADCs). CDS and fowler sampling are done in the software to reduce the offset voltage into random noise. This paper focuses on the former two subsystems, which mainly cause the time-variation components among the three subsystems. The parameters and symbols used for the mathematical analysis in Section \ref{sec:analysis} are compiled in the table of Appendix \ref{sec:appendixA}.

\subsection{Formulation of output voltage} \label{subsec:formulation}

\begin{figure}[H]
\begin{center}
\includegraphics[width=15cm]{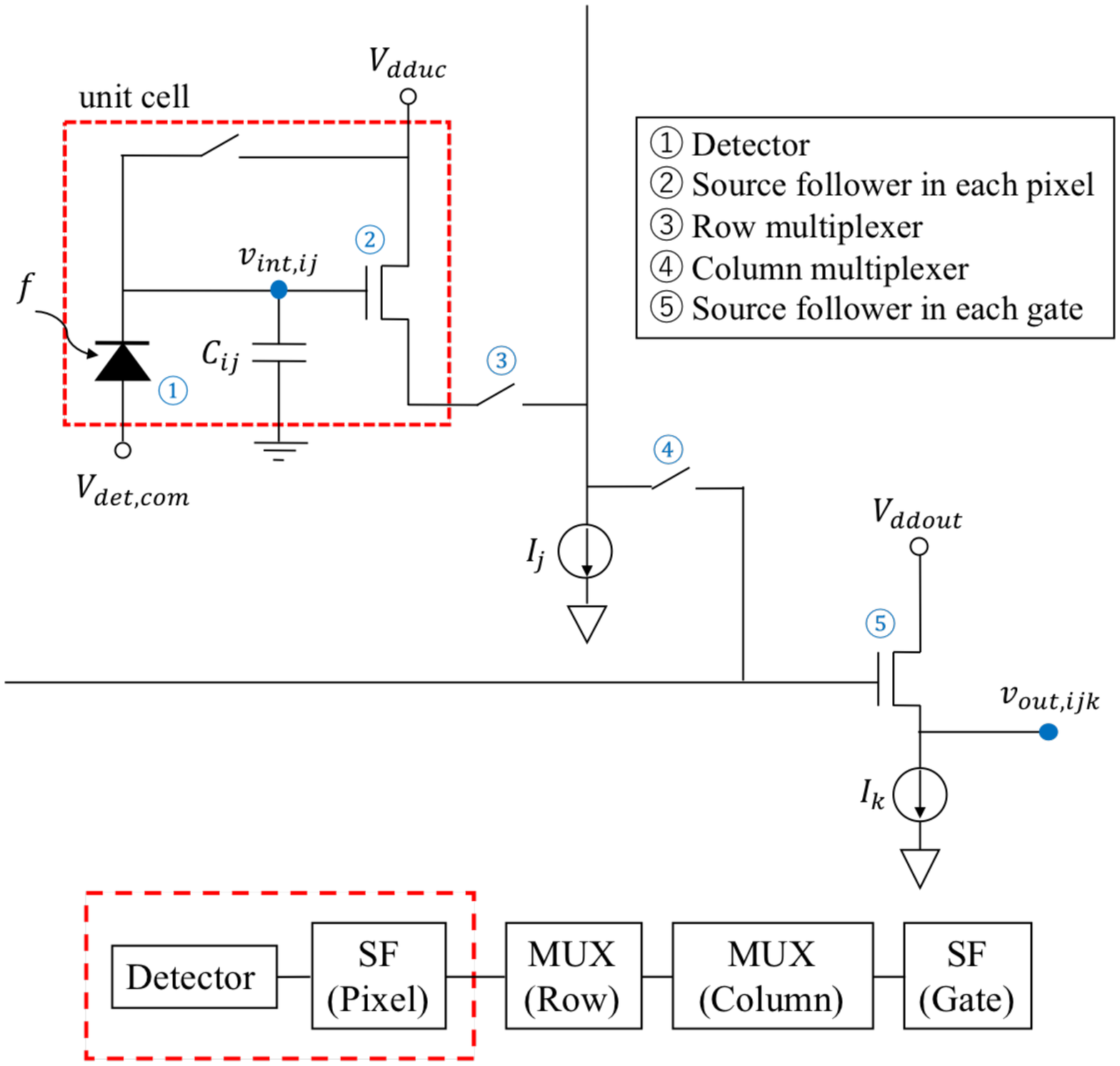}
\caption{Equivalent circuit of an infrared detector system (top) and flowchart of its signal chain (bottom). \label{fig:equivalent_circuit}}
\end{center}
\end{figure}

We first formulate the output voltage at the k-th readout gate of the ROIC in the infrared detector. Figure \ref{fig:equivalent_circuit} shows an equivalent circuit of the ROIC \citep{2009ITED...56.2506J} and the signal flow from receiving photons on the detector plane to sending voltages converted from photoelectrons to the warm electronics. The signal flow from the semiconductor device to the gate of the ROIC is mainly divided into three. First, the photoelectrons and electrons corresponding to dark current are converted into voltage in each pixel unit through the charging of the electrons at its condenser. Note that the electrons are integrated at the gate of the source follower in each unit cell for the MIRI detector developed for the JWST. When the number of photons per second $f_{ij}$ at the observing wavelength $\lambda$ falls on the (i, j) detector pixel and the generated electrons are accumulated at the condenser during the exposure time $t_{exp}$, the output voltage at the gate of the source follower in each unit cell $v_{int,ij}$, is written as
\begin{equation}
\label{equ:v_int}
v_{int,ij}(\lambda,t) \approx \frac{\alpha_{ij}\eta_{ij}f_{ij}(\lambda,t)+I_{dark,ij}}{C_{ij}}t_{exp}+V_{res,ij} ,
\end{equation}
where $C_{ij}$ is the capacitor capacitance of the (i, j) pixel, $I_{dark,ij}$ is the dark current in the (i, j) pixel and $V_{res,ij}$ is the offset voltage generated when resetting the electrons accumulated at the condenser of the (i, j) pixel. Next, the voltage is transferred to the column multiplexer through the source follower of each pixel and to the source follower at each readout gate of the ROIC. The output voltage at the readout gate $v_{out,ijk}$, is
\begin{equation}
v_{out,ijk}(\lambda,t) \approx A_{k}(t)A_{ij}(t)\gamma_{ij}\{\alpha_{ij}\eta_{ij}f_{ij}(\lambda,t)+I_{dark,ij}\} ,
\end{equation}
where $A_{ij}$ and $A_k$ are the transfer function of the source follower in the (i, j) pixel and at the k-th readout gate, respectively, and $\gamma_{ij} \equiv t_{exp}/C_{ij}$. Note that, although $v_{out}$ is generally calculated as a convolution of $A_{ij}$, $A_{k}$, and $\gamma_{ij}\{\alpha_{ij}\eta_{ij}f_{ij}(\lambda,t)+I_{dark,ij}\}$, the convolution is approximately replaced by multiplication because the source followers applied to astronomical detectors have wide bandwidth \citep{1999ITED...46...96J, 2009ITED...56.2506J}. $V_{res,ij}$ present in Equation (\ref{equ:v_int}) can be reduced to a random component with CDS or fowler sampling in the software. Thus, the transit signal is mostly affected by the time-variations of the two effective gains multiplied with $f_{ij}(\lambda,t)$. The offset voltage, however, is present in the multiplexer of ROIC; its impact on the transit signal is evaluated in Section \ref{sec:discussion}.

\subsection{Output voltages for three types of pixels} \label{subsec:three}

We apply the densified pupil spectrograph to the infrared detector system and then formulate the output voltages at the k-th readout gate for the three types of the pixels: the science, background, and reference pixels. In analogy with the mathematical analysis introduced in Section \ref{subsec:method}, the voltage at the k-th readout gate of ROIC is decomposed into the sum of the average, fluctuation, and random terms of $v_{out,ijk}$:
\begin{equation}
\label{equ:v_out,ijk}
v_{out,ijk}(\lambda,t) = \hoge<v_{out,ijk}(\lambda,t)>_{t}+{\delta}v_{out,ijk}(\lambda,t)+\sigma_{out,ijk}(\lambda) ,
\end{equation}
where $\hoge<v_{out,ijk}(\lambda,t)>_t$ is the time-average of $v_{out,ijk}(\lambda,t)$ over one transit observation, ${\delta}v_{out,ijk}(\lambda,t)$ is the time-variation of $v_{out,ijk}(\lambda,t)$ and $\sigma_{out,ijk}(\lambda)$ is the random components associated to $v_{out,ijk}(\lambda,t)$. The time-averaged values of the science, background, and reference pixels, $\hoge<v_{sci,ijk}(\lambda,t)>_t$, $\hoge<v_{back,ijk}(\lambda,t)>_t$, and $\hoge<v_{ref,ijk}(t)>_t$, are
\begin{equation}
\label{equ:v_sci,ijk}
\hoge<v_{sci,ijk}(\lambda,t)>_{t} \approx \hoge<A_{k}(t)>_t\hoge<A_{ij}(t)>_{t}\gamma_{ij}[\alpha_{ij}\eta_{ij}\{\hoge<f_{s}(\lambda,t)>_{t}+f_{z}(\lambda)\}+I_{dark,ij}] ,
\end{equation}
\begin{equation}
\hoge<v_{back,ijk}(\lambda,t)>_{t} \approx \hoge<A_{k}(t)>_t\hoge<A_{ij}(t)>_{t}\gamma_{ij}\{\alpha_{ij}\eta_{ij}f_{z}(\lambda)+I_{dark,ij}\} ,
\end{equation}
and
\begin{equation}
\hoge<v_{ref,ijk}(t)>_{t} \approx \hoge<A_{k}(t)>_t\hoge<A_{ij}(t)>_{t}\gamma_{ij}I_{dark,ij} ,
\end{equation}
where $f_s$ and $f_z$ are the numbers of photons from a target object and the zodiacal light falling on a detector pixel, respectively. Given that more than second-order fluctuations are ignored and the variation of the zodiacal light ${\delta}f_z$, is much smaller than the transit signal ${\delta}f_s$, the time-variation terms in Equation (\ref{equ:v_out,ijk}) for the three types of the pixels, ${\delta}v_{sci,ijk}$, ${\delta}v_{back,ijk}$, and ${\delta}v_{ref,ijk}$ become
\begin{eqnarray}
\label{equ:dv_sci}
{\delta}v_{sci,ijk}(\lambda,t) &\approx& \hoge<A_{k}(t)>_t\hoge<A_{ij}(t)>_{t}\gamma_{ij}\alpha_{ij}\eta_{ij}{\delta}f_{s}(\lambda,t) \nonumber \\
&+&\hoge<A_{ij}(t)>_{t}\gamma_{ij}[\alpha_{ij}\eta_{ij}\{\hoge<f_{s}(\lambda,t)>_{t}+f_{z}(\lambda)\}+I_{dark,ij}]{\delta}A_{k}(t) \nonumber \\
&+&\hoge<A_{k}(t)>_{t}\gamma_{ij}[\alpha_{ij}\eta_{ij}\{\hoge<f_{s}(\lambda,t)>_{t}+f_{z}(\lambda)\}+I_{dark,ij}]{\delta}A_{ij}(t) ,
\end{eqnarray}
\begin{eqnarray}
\label{equ:dv_back}
{\delta}v_{back,ijk}(\lambda,t) &\approx& \hoge<A_{ij}(t)>_{t}\gamma_{ij}\{\alpha_{ij}\eta_{ij}f_{z}(\lambda)+I_{dark,ij}\}{\delta}A_{k}(t) \nonumber \\
&+&\hoge<A_{k}(t)>_{t}\gamma_{ij}\{\alpha_{ij}\eta_{ij}f_{z}(\lambda)+I_{dark,ij}\}{\delta}A_{ij}(t) ,
\end{eqnarray}
and
\begin{equation}
\label{equ:dv_ref}
{\delta}v_{ref,ijk}(t) \approx \hoge<A_{ij}(t)>_{t}\gamma_{ij}I_{dark,ij}{\delta}A_{k}(t)+\hoge<A_{k}(t)>_{t}\gamma_{ij}I_{dark,ij}{\delta}A_{ij}(t) ,
\end{equation}
The terms of ${\delta}A_{ij}$ and ${\delta}A_{k}$ in Equations (\ref{equ:dv_sci}) to (\ref{equ:dv_ref}) correspond to the systematic components that have an impact on the transit light curve. Considering each random component is statistically independent, the random noises acting on the three types of the pixels, $\sigma_{sci,ijk}$, $\sigma_{back,ijk}$, and $\sigma_{ref,ijk}$, can be written as
\begin{equation}
\label{equ:sig_sci}
\sigma_{sci,ijk}(\lambda) \approx \sqrt{\hoge<A_{k}(t)>^2_t\hoge<A_{ij}(t)>^2_{t}\gamma^2_{ij}[\alpha^2_{ij}\eta^2_{ij}\{\sigma^2_{photon,s}(\lambda)+\sigma^2_{photon,z}(\lambda)\}+\sigma^2_{dark,ij}]+\sigma^2_{read,ijk}} ,
\end{equation}
\begin{equation}
\label{equ:sig_back}
\sigma_{back,ijk}(\lambda) \approx \sqrt{\hoge<A_{k}(t)>^2_t\hoge<A_{ij}(t)>^2_{t}\gamma^2_{ij}\{\alpha^2_{ij}\eta^2_{ij}\sigma^2_{photon,z}(\lambda)+\sigma^2_{dark,ij}\}+\sigma^2_{read,ijk}} ,
\end{equation}
and
\begin{equation}
\label{equ:sig_ref}
\sigma_{ref,ijk} \approx \sqrt{\hoge<A_{k}(t)>^2_t\hoge<A_{ij}(t)>^2_{t}\gamma^2_{ij}\sigma^2_{dark,ij}+\sigma^2_{read,ijk}} ,
\end{equation}
where $\sigma_{photon,s}$, $\sigma_{photon,z}$, and $\sigma_{dark}$ are the shot noises of an object light, zodiacal light, and dark current, respectively.

\subsection{Calibration process} \label{subsec:calibration}

We optimize the calibration method introduced in Section \ref{subsec:method} for the infrared detector systems and mathematically show how to calibrate the time-variation components with the three types of pixels. The process of the calibration method is mainly divided into four. The first process of the calibration is to average the pixel values over all the pixels exposed by the same spectrally-resolved light for the science and background pixels $v_{sci,ijk}(\lambda,t)$ and $v_{back,ijk}(\lambda,t)$, as well as for those over all the reference pixels, $v_{ref,ijk}(t)$, to smooth out pixel-to-pixel time-variations and extract common systematic components of the entire pixels. The pixel-averaged science and background values are
\begin{equation}
\label{equ:v_sci_ave}
\hoge<v_{sci,ij}(\lambda,t)>_{pix} = \frac{\Sigma_{ijk}v_{sci,ijk}(\lambda,t)}{n_{sci}(\lambda)} = \frac{\Sigma_{ijk}\hoge<v_{sci,ijk}(\lambda,t)>_{t}}{n_{sci}(\lambda)}+\frac{\Sigma_{ijk}{\delta}v_{sci,ijk}(\lambda,t)}{n_{sci}(\lambda)}+\frac{\sqrt{\Sigma_{ijk}\sigma^2_{sci,ijk}(\lambda)}}{n_{sci}(\lambda)} ,
\end{equation}
\begin{equation}
\hoge<v_{back,ij}(\lambda,t)>_{pix} = \frac{\Sigma_{ijk}v_{back,ijk}(\lambda,t)}{n_{back}(\lambda)} = \frac{\Sigma_{ijk}\hoge<v_{back,ijk}(\lambda,t)>_{t}}{n_{back}(\lambda)}+\frac{\Sigma_{ijk}{\delta}v_{back,ijk}(\lambda,t)}{n_{back}(\lambda)}+\frac{\sqrt{\Sigma_{ijk}\sigma^2_{back,ijk}(\lambda)}}{n_{back}(\lambda)} ,
\end{equation}
where $n_{sci}(\lambda)$ and $n_{back}(\lambda)$ are the numbers of the science and background pixels exposed by the same component of $\lambda$. The pixel-averaged value over the reference pixels becomes
\begin{equation}
\label{equ:v_ref_ave}
\hoge<v_{ref,ij}(t)>_{pix} = \frac{\Sigma_{ijk}v_{ref,ijk}(t)}{n_{sci}} = \frac{\Sigma_{ijk}\hoge<v_{ref,ijk}(t)>_{t}}{n_{ref}}+\frac{\Sigma_{ijk}{\delta}v_{ref,ijk}(t)}{n_{ref}}+\frac{\sqrt{\Sigma_{ijk}\sigma^2_{ref,ijk}}}{n_{ref}} ,
\end{equation}
where $n_{ref}$ is the number of all the reference pixels. Using Equations of (\ref{equ:v_sci,ijk}) to (\ref{equ:dv_ref}), Equations of (\ref{equ:v_sci_ave}) to (\ref{equ:v_ref_ave}) become
\begin{eqnarray}
\label{equ:<v_sci,ij>}
\hoge<v_{sci,ij}(\lambda,t)>_{pix} &=& \hoge<A_{com}(t)>_{t}\hoge<\gamma_{ij}>_{pix}[\hoge<\alpha_{ij}\eta_{ij}>_{pix}\{\hoge<f_{s}(\lambda,t)>_{t}+f_{z}(\lambda)\}+\hoge<I_{dark,ij}>_{pix}] \nonumber \\
&+& \hoge<A_{com}(t)>_{t}\hoge<\gamma_{ij}>_{pix}\hoge<\alpha_{ij}\eta_{ij}>_{pix}{\delta}f_{s}(\lambda,t) \nonumber \\
&+& \hoge<\gamma_{ij}>_{pix}[\hoge<\alpha_{ij}\eta_{ij}>_{pix}\{\hoge<f_{s}(\lambda,t)>_{t}+f_{z}(\lambda)\}+\hoge<I_{dark,ij}>_{pix}]{\delta}A_{com}(t)+\sqrt{\hoge<\sigma^{2}_{sci,ij}(\lambda)>_{pix}} \nonumber \\
&+& \sqrt{\hoge<\sigma^{2}_{sci,ij}(\lambda)>_{pix}}\frac{{\delta}A_{com}(t)}{\hoge<A_{com}(t)>_{t}} , \\
\hoge<v_{back,ij}(\lambda,t)>_{pix} &=& \hoge<A_{com}(t)>_{t}\hoge<\gamma_{ij}>_{pix}\{\hoge<\alpha_{ij}\eta_{ij}>_{pix}f_{z}(\lambda)+\hoge<I_{dark,ij}>_{pix}\} \nonumber \\
&+& \hoge<\gamma_{ij}>_{pix}\{\hoge<\alpha_{ij}\eta_{ij}>_{pix}f_{z}(\lambda)+\hoge<I_{dark,ij}>_{pix}\}{\delta}A_{com}(t)+\sqrt{\hoge<\sigma^{2}_{back,ij}(\lambda)>_{pix}} \nonumber \\
&+& \sqrt{\hoge<\sigma^{2}_{back,ij}(\lambda)>_{pix}}\frac{{\delta}A_{com}(t)}{\hoge<A_{com}(t)>_{t}} , \\
\label{equ:<v_ref,ij>}
\hoge<v_{ref,ij}(t)>_{pix} &=& \hoge<A_{com}(t)>_{t}\hoge<\gamma_{ij}>_{pix}\hoge<I_{dark,ij}>_{pix}+\hoge<\gamma_{ij}>_{pix}\hoge<I_{dark,ij}>_{pix}{\delta}A_{com}(t)+\sqrt{\hoge<\sigma^{2}_{ref,ij}>_{pix}} \nonumber \\
&+& \sqrt{\hoge<\sigma^{2}_{ref,ij}>_{pix}}\frac{{\delta}A_{com}(t)}{\hoge<A_{com}(t)>_{t}} ,
\end{eqnarray}
The second process is to remove the time-variation components from the pixel-averaged science data using the pixel-averaged background and reference data. Focusing on a fact that the time-variation components associated to the background and reference data are composed of only the systematic components present in the semiconductor device and ROIC, as shown in Equations (\ref{equ:dv_back}) and (\ref{equ:dv_ref}), the systematic components are derived by subtraction of the time- and pixel-averaged background and reference data over one transit observation, $\hoge<v_{back,ij}(\lambda,t)>_{pix,t}$, $\hoge<v_{ref,ij}(t)>_{pix,t}$, from the pixel-averaged background and reference data, $\hoge<v_{back,ij}(\lambda,t)>_{pix}$, $\hoge<v_{ref,ij}(t)>_{pix}$, at each frame. Note that, because the amplitude of the time-variation component is proportional to the signal amplitude, the time-variation component associated to the science pixel can be reconstructed through multiplication of $\{\hoge<v_{back,ij}(\lambda,t)>_{pix}+\hoge<v_{ref,ij}(t)>_{pix}\}-\{\hoge<v_{back,ij}(\lambda,t)>_{pix,t}+\hoge<v_{ref,ij}(t)>_{pix,t}\}$ with the ratio of the time- and pixel-averaged science data to the sum of the background and reference data, $\frac{\hoge<v_{sci,ij}(\lambda,t)>_{pix,t}}{\hoge<v_{back,ij}(\lambda,t)>_{pix,t}+\hoge<v_{ref,ij}(t)>_{pix,t}}$. Using Equations of (\ref{equ:<v_sci,ij>}) to (\ref{equ:<v_ref,ij>}), the systematic component can be removed from the science data as follows:
\begin{eqnarray}
\label{equ:v_cal}
v_{cal}(\lambda,t) &=& \hoge<v_{sci,ij}(\lambda,t)>_{pix}-[\{\hoge<v_{back,ij}(\lambda,t)>_{pix}+\hoge<v_{ref,ij}(t)>_{pix}\}-\{\hoge<v_{back,ij}(\lambda,t)>_{pix,t}+\hoge<v_{ref,ij}(t)>_{pix,t}\}] \nonumber \\
&\times& \frac{\hoge<v_{sci,ij}(\lambda,t)>_{pix,t}}{\hoge<v_{back,ij}(\lambda,t)>_{pix,t}+\hoge<v_{ref,ij}(t)>_{pix,t}} \nonumber \\
&\approx& \hoge<v_{sci,ij}(\lambda,t)>_{pix,t}+\hoge<A_{com}(t)>_{t}\hoge<\gamma_{ij}>_{pix}\hoge<\alpha_{ij}\eta_{ij}>_{pix}{\delta}f_{s}(\lambda,t)+\hoge<v_{sci,ij}(\lambda,t)>_{pix,t}\frac{{\delta}A_{com}(t)}{\hoge<A_{com}(t)>_{t}} \nonumber \\
&+& \sqrt{\hoge<\sigma^{2}_{sci,ij}(\lambda)>_{pix}}+\sqrt{\hoge<\sigma^{2}_{sci,ij}(\lambda)>_{pix}}\frac{{\delta}A_{com}(t)}{\hoge<A_{com}(t)>_{t}} \nonumber \\
&-& \left[\{\hoge<v_{back,ij}(\lambda,t)>_{pix,t}+\hoge<v_{ref,ij}(t)>_{pix,t}\}\frac{{\delta}A_{com}(t)}{\hoge<A_{com}(t)>_{t}}+\sqrt{\hoge<\sigma^{2}_{back,ij}(\lambda)>_{pix}+\hoge<\sigma^{2}_{ref,ij}>_{pix}}\left\{1+\frac{{\delta}A_{com}(t)}{\hoge<A_{com}(t)>_{t}}\right\}\right] \nonumber \\
&\times& \frac{\hoge<v_{sci,ij}(\lambda,t)>_{pix,t}}{\hoge<v_{back,ij}(\lambda,t)>_{pix,t}+\hoge<v_{ref,ij}(t)>_{pix,t}} \nonumber \\
&\approx& \hoge<A_{com}(t)>_{t}\hoge<\gamma_{ij}>_{pix}[\hoge<\alpha_{ij}\eta_{ij}>_{pix}\{\hoge<f_{s}(\lambda,t)>_{t}+f_{z}(\lambda)\}+\hoge<I_{dark,ij}>_{pix}] \nonumber \\
&+& \hoge<A_{com}(t)>_{t}\hoge<\gamma_{ij}>_{pix}\hoge<\alpha_{ij}\eta_{ij}>_{pix}{\delta}f_{s}(\lambda,t)+\sigma_{cal}(\lambda)+\sigma_{cal}(\lambda)\frac{{\delta}A_{com}(t)}{\hoge<A_{com}(t)>_{t}} ,
\end{eqnarray}
\begin{equation}
\sigma_{cal}(\lambda) = \sqrt{\hoge<\sigma^{2}_{sci,ij}(\lambda)>_{pix}+\left\{\frac{\hoge<v_{sci,ij}(\lambda,t)>_{pix,t}}{\hoge<v_{back,ij}(\lambda,t)>_{pix,t}+\hoge<v_{ref,ij}(t)>_{pix,t}}\right\}^2\{\hoge<\sigma^{2}_{back,ij}(\lambda)>_{pix}+\hoge<\sigma^{2}_{ref,ij}>_{pix}\}} ,
\end{equation}
where $v_{cal}(\lambda,t)$ and $\sigma_{cal}(\lambda)$ are the calibrated science data and the random term associated to $v_{cal}(\lambda,t)$ after the second process, respectively. Note that the common systematic component can be reduced to a random one when the gain is perfectly proportional to various signal levels. The next process is to remove the offset component included in the science data reduced through the second process. Since the sum of the zodiacal light and the dark current is equal to the time- and pixel-averaged value of the background pixels, the offset components can be removed through subtraction of $\hoge<v_{back,ij}(\lambda,t)>_{pix,t}$ from $v_{cal}(\lambda,t)$:
\begin{eqnarray}
\label{equ:v_sub}
v_{sub}(\lambda,t) &\approx& v_{cal}(\lambda,t)-\hoge<v_{back,ij}(\lambda,t)>_{pix,t} \nonumber \\
&=& \hoge<A_{com}(t)>_{t}\hoge<\gamma_{ij}>_{pix}[\hoge<\alpha_{ij}\eta_{ij}>_{pix}\{\hoge<f_{s}(\lambda,t)>_{t}+f_{z}(\lambda)\}+\hoge<I_{dark,ij}>_{pix}] \nonumber \\
&+& \hoge<A_{com}(t)>_{t}\hoge<\gamma_{ij}>_{pix}\hoge<\alpha_{ij}\eta_{ij}>_{pix}{\delta}f_{s}(\lambda,t)+\sigma_{cal}(\lambda)+\sigma_{cal}(\lambda)\frac{{\delta}A_{com}(t)}{\hoge<A_{com}(t)>_{t}} \nonumber \\
&-& \hoge<A_{com}(t)>_{t}\hoge<\gamma_{ij}>_{pix}\{\hoge<\alpha_{ij}\eta_{ij}>_{pix}f_{z}(\lambda)+\hoge<I_{dark,ij}>_{pix}\} \nonumber \\
&=& \hoge<A_{com}(t)>_{t}\hoge<\gamma_{ij}>_{pix}\hoge<\alpha_{ij}\eta_{ij}>_{pix}\hoge<f_{s}(\lambda,t)>_{t}+\hoge<A_{com}(t)>_{t}\hoge<\gamma_{ij}>_{pix}\hoge<\alpha_{ij}\eta_{ij}>_{pix}{\delta}f_{s}(\lambda,t)+\sigma_{cal}(\lambda) \nonumber \\
&+& \sigma_{cal}(\lambda)\frac{{\delta}A_{com}(t)}{\hoge<A_{com}(t)>_{t}} .
\end{eqnarray}
Here, the noise introduced during the subtraction is considered as negligible because the residual offset is not expected to change over a single transit observation. Last, we can obtain the normalized signal, $v_{sub,norm}$, using $v_{sub}(\lambda,t)$ of Equation (\ref{equ:v_sub}):
\begin{equation}
\label{equ:v_sub,norm}
v_{sub,norm}(\lambda,t) \approx 1+\frac{{\delta}f_{s}(\lambda,t)}{\hoge<f_{s}(\lambda,t)>_{t}}+\sigma_{sub,norm}(\lambda)+\Delta_{sub,norm}(\lambda,t) ,
\end{equation}
where $\sigma_{sub,norm}(\lambda)$ and $\Delta_{sub,norm}(\lambda,t)$ are the random and the quadratic terms associated to $v_{sub,norm}(\lambda,t)$:
\begin{equation}
\label{equ:sigma_sub,norm}
\sigma_{sub,norm}(\lambda) = \frac{\sigma_{cal}(\lambda)}{\hoge<A_{com}(t)>_{t}\hoge<\gamma_{ij}>_{pix}\hoge<\alpha_{ij}\eta_{ij}>_{pix}\hoge<f_{s}(\lambda,t)>_{t}} ,
\end{equation}
\begin{equation}
\label{equ:Delta_sub,norm}
\Delta_{sub,norm}(\lambda,t) = \frac{\sigma_{cal}(\lambda)}{\hoge<A_{com}(t)>_{t}\hoge<\gamma_{ij}>_{pix}\hoge<\alpha_{ij}\eta_{ij}>_{pix}\hoge<f_{s}(\lambda,t)>_{t}}\frac{{\delta}A_{com}(t)}{\hoge<A_{com}(t)>_{t}} = \sigma_{sub,norm}(\lambda)\frac{{\delta}A_{com}(t)}{\hoge<A_{com}(t)>_{t}} .
\end{equation}
As shown in Equation (\ref{equ:Delta_sub,norm}), when the gain variation, $\frac{{\delta}A_{com}}{\hoge<A_{com}>_t}$, is small, the quadratic term, $\Delta_{sub,norm}$ can be ignored. Thus, the systematic component common in the entire detector plane can be almost reduced to the random noise; however, the random noise obtained through the calibration process is larger than the one acting on the original science data, as shown in Equation (\ref{equ:sigma_sub,norm}). As the integration time increases, $\sigma_{sub,norm}$ further decreases and the transit signal can be more accurately measured.

\section{Simulation} \label{sec:simulation}

In this section, we perform numerical simulations to investigate how much the spectro-photometric stability of a densified pupil spectrograph optimized for the mid-infrared wavelength range can be improved with the proposed calibration method under a future large cryogenic telescope such as the Origins Space Telescope \citep{2018arXiv180307730F}.

\subsection{Setup} \label{subsec:setup}

We first introduce the appropriate assumptions to perform valid evaluations on the effectiveness of the calibration technique. Four cases were considered for the numerical simulations. The distances to all the target systems were fixed to 10 pc. The effective temperatures of the four host stars were set to 2,500, 3,000, 3,500, and 4,000 K. The stellar radii were 0.1, 0.16, 0.39, and 0.60 times the Sun radius, respectively. The four host stars were approximately corresponding to late-M, mid-M, early-M, and late-K type stars. Note that, since the limb darkening has a small impact on the transit light curve in the mid-infrared wavelength range, no limb darkening was assumed to exist in the numerical simulations. The transiting planet orbiting each host star was assumed to be a terrestrial planet with radius of 1 Earth radius and effective temperature of 288 K. Given that the albedos of the target planets were 0.306 and the inclinations were fixed to 90 degrees, the semi-major axes of the planets were set to 0.015, 0.030, 0.110, and 0.225 AU for the four simulation cases, respectively; the orbital periods were 5.2, 6.8, 21.1, and 48.9 days. The light from the target system was embedded in the zodiacal light that was almost blackbody with an effective temperature of 275 K. Based on the previous observation \citep{2016AJ....151...71K}, the surface brightness of the zodiacal light was set to 5 MJy/sr at 9 \textmu m. The characteristics of the four target systems are compiled in Table \ref{tab:target}.

\begin{figure}[H]
\begin{center}
\includegraphics[width=11cm]{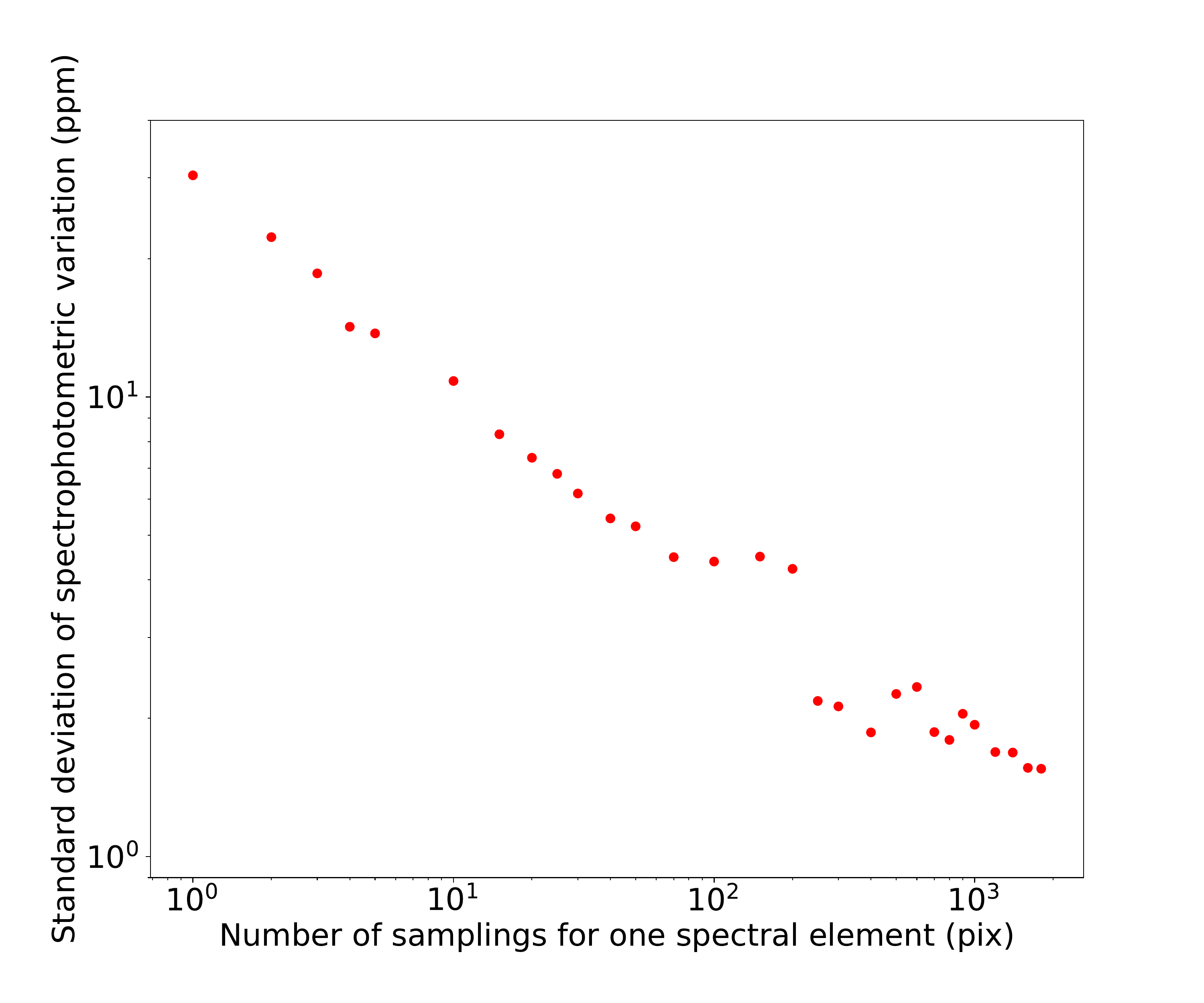}
\caption{Standard deviation of spectrophotometric variation over the entire wavelength range of 3-22um as a function of the number of the samplings for each spectral element. The effect of the uncertainty is reduced with being proportion to the inverse of the square root of the number of pixels. Since the number of the samplings is set to about 2,000 pixels for this simulation, the flat-field uncertainty is negligible. \label{fig:flat_field}}
\end{center}
\end{figure}

\begin{figure}[H]
\begin{center}
\includegraphics[width=14cm]{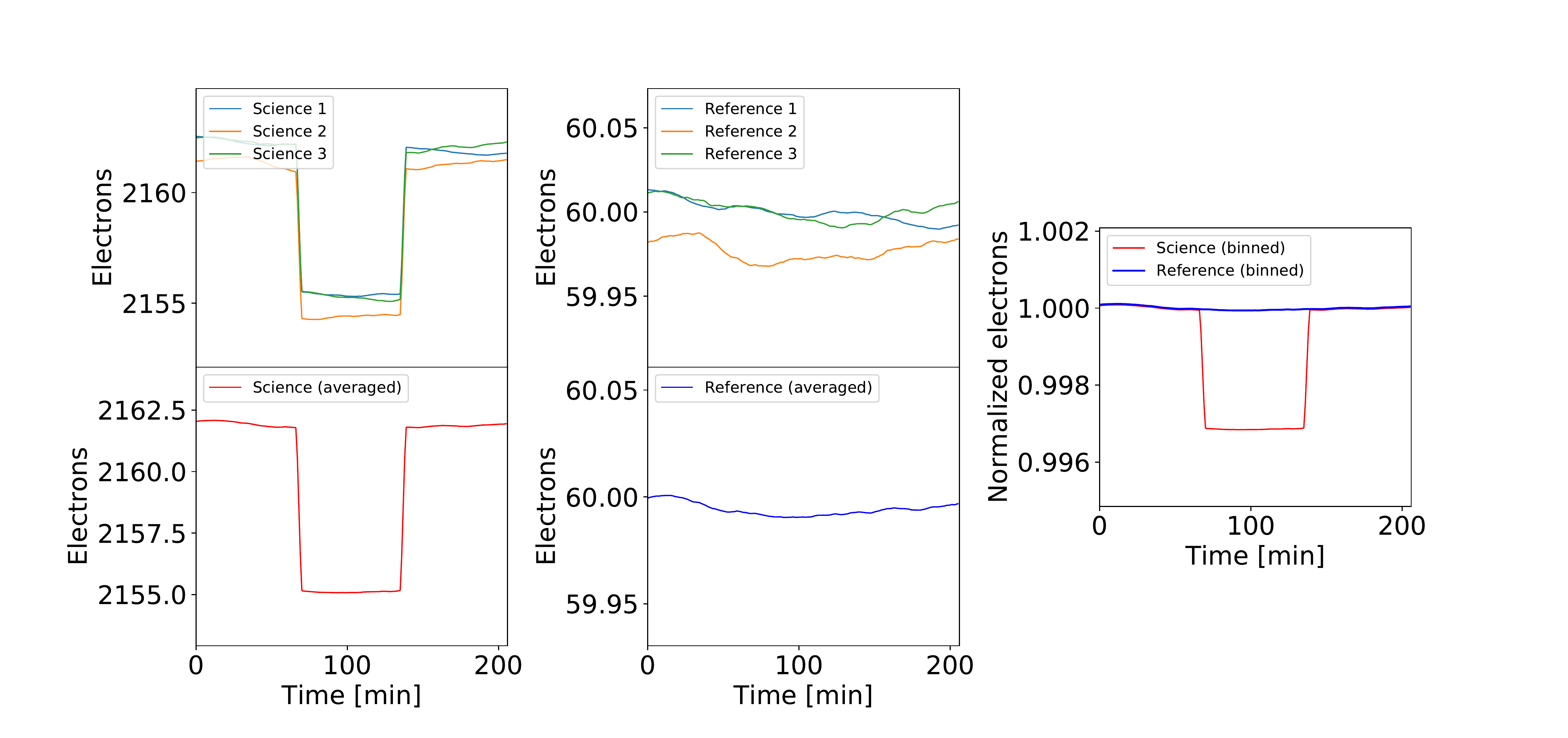}
\caption{Comparison of the science pixel data with the reference pixel one without the shot noise. The light blue, orange, and green lines in the upper left and right figures show three example data for the science and reference pixels. The lower left and right figures are the average data over the science and reference pixels, respectively. The pixel-to-pixel time-variations of each data are almost smoothed out. The right figure compares the average science data (red solid line) with the average reference one (blue solid line). The averaged reference data accurately traces the time-variation components associated to the science data.\label{fig:smoothed_out}}
\end{center}
\end{figure}

The target systems were assumed to be observed with a large cryogenic space telescope such as the Origins Space Telescope. The primary mirror of the Origins Space Telescope is 9.24 m in diameter. The densified pupil spectrograph was applied to the transit observation. We assumed that the systematic components generated due to wavefront aberration in the telescope and instrument systems can be ignored because the system provides multiple spectra of the divided primary mirror on the detector plane. The transit instrument applies two mercury-cadmium-telluride (MCT) detectors \citep[e.g.,][]{2013SPIE.1306..6978} and one Si:As blocked-impurity-band (BIB) detector \citep[e.g.,][]{2015PASP..127..665R, 2015PASP..127..675R} to cover the wide wavelength range of 3 to 11 and 11 to 22 \textmu m, respectively. The wavelength ranges of the spectra formed on the three detectors are 3-6, 6-11, and 11-22 \textmu m, respectively. The pixel formats of the three detectors are 1,024$\times$1,024. The numbers of the science, background, and reference pixels for each detector were set to 120,000, 120,000, and 760,000, respectively. The field of view of the field stop used for this simulation was set to 2 arcsec in radius, such that the photometric variation caused by the PSF motion loss on the field stop under a telescopic jitter of 10 mas was smaller than 1 ppm. Note that the PSF motion loss on the field stop can be calculated based on a mathematical formula introduced by \cite{2017AJ....154...97I}. The optical throughput of the instrument, including the quantum efficiency of the detector, is 30 \% for each band. The minimum spectral-resolved wavelength, ${\Delta}{\lambda}$, was set to 0.01 times the central wavelength of each spectral channel, 0.045 \textmu m for the shorter band, 0.085 \textmu m for the middle one, and 0.165 \textmu m for the longer one, almost corresponding to the spectral resolution of 100 for the general-purpose spectrographs. Note that, while general-purpose spectrographs provide a constant ${\lambda}/{\Delta}{\lambda}$, ${\Delta}{\lambda}$ was fixed for the densified pupil spectral graph. The dark current of each detector was set to 1.0 e-/sec for each band. The readout noise was set to approximately 5.5 e- per read for both the MCT and Si:As detectors, assuming that CDS sampling and a fowler-16 sampling were applied to the MCT and Si:As detectors, respectively \citep[e.g.,][]{2014PASP..126..739R, 2015PASP..127.1144R}. The exposure time of each frame was set to 60 seconds for each simulation case, which is long enough that the dark current noise dominates the readout noise at one exposure for the reference pixels. Based on the flat-field-uncertainty map derived from the WISE W3 data collected during the WISE cryogenic mission, the corrected effective gain after the flat-field calibration, corresponding to the conversion factor from electron to voltage, ranged from 0.99985 to 1.00015. Because the background and science spectra are focused at different locations on the detector plane, as shown in Figure \ref{fig:detector_plane}, the subtraction of the background component from the object one cannot perfectly remove the offsets on the science data due to the flat-field uncertainty. Figure \ref{fig:flat_field} shows the standard deviation of the systematic errors over the shorter wavelength of 3 to 22 \textmu m generated by the subtraction of the background light as a function of the number of the pixels for each spectral element. The flat-field uncertainty map of the WISE W3 data was used for this evaluation. As shown in Figure \ref{fig:flat_field}, the impact of the flat-field uncertainty on this method is negligible, owing to the number of samplings for each spectral element. We also investigated whether the pixel-to-pixel time-variation can be smoothed out through averaging the data. Figure \ref{fig:smoothed_out} compares the science data with the reference set before and after the average without shot noise. The pixel-to-pixel time-variation components were almost smoothed out through average and the common components over the entire detector were extracted. Therefore, we assumed that the common components over the entire detector, $A_{com}$, remain in the transit light curve for this simulation. The standard deviation of the common component of the gain, $\sigma_{A_{com}}$, is set as follows:
\begin{equation}
\sigma_{A_{com}} \equiv \sqrt{\int_{f_{min}}^{f_{max}} \sigma^2_{g}(f) df}=1.0\times10^{-4} ,
\end{equation}
where $\sigma_{g}$ is the standard deviation of the common component as a function of frequency, $f$, and is in inverse proportion to the root mean square of the frequency:
\begin{equation}
\sigma_{g}(f) \approx \frac{4.1\times10^{-5}}{\sqrt{f}} .
\end{equation}
The frequencies of the common time-variation components range from $2\times10^{-5} (=f_{min})$ to $8\times10^{-3} (=f_{max})$ Hz. Based on the previous Spitzer observations (Knutson et al. 2009; Knutson et al. 2011), the standard deviation of the time-variation components was set to 100 ppm. Note that the effective gain applied at the four readout gates ($k$=1,2,3 and 4) $A_{k}$, also remains because of the limited number of the readout gates, which are four different effective gains, applied in this simulation. The offset estimation error is also negligible. The parameters of the telescope and the instrument are compiled in Table \ref{tab:ost}.

\begin{deluxetable}{lcccc}[H]
\tablecaption{Characteristics of the target system \label{tab:target}}
\tablehead{Items & Case 1 & Case 2 & Case 3 & Case 4}
\startdata
\bf{Star} & \multicolumn{4}{c}{\bf{-}} \\
Distance (pc) & 10 & 10 & 10 & 10 \\
Effective temperature (K) & 2,500 & 3,000 & 3,500 & 4,000 \\
Radius ($\rm R_\odot$)\tablenotemark{*} & 0.10 & 0.16 & 0.39 & 0.60 \\
\hline
\bf{Planet} & \multicolumn{4}{c}{\bf{Terrestrial planet}} \\
Inclination (deg) & 90 & 90 & 90 & 90 \\
Albedo & 0.31 & 0.31 & 0.31 & 0.31 \\
Semi-major axis ($10^{-3}$ AU) & 14.6 & 33.0 & 110.3 & 224.6 \\
Period (days) & 5.2 & 6.8 & 21.1 & 48.9 \\
Equilibrium temperature (K) & 288.2 & 288.2 & 288.2 & 288.2 \\
Radius ($\rm R_\oplus$) & 1 & 1 & 1 & 1 \\
\hline
\bf{Zodiacal light} & \multicolumn{4}{c}{\bf{-}} \\
Level at 9 \textmu m (MJy/sr) & 5 & 5 & 5 & 5 \\
Effective temperature (K) & 275 & 275 & 275 & 275 \\
Spectrum & Black body & Black body & Black body & Black body
\enddata
\tablenotetext{*}{We cited \cite{2012ApJ...757..112B}, \cite{2017ApJ...845..178B} and \cite{2015ApJ...804...64M} for the stellar radii of the cases 2, 3 and 4. On the other hand, we referred to the stellar parameters of trapppist-1 system \citep{2017Natur.542..456G, 2018ApJ...853...30V} for the stellar radius of the case 1.}
\end{deluxetable}

\begin{deluxetable}{lccc}[H]
\tablecaption{Specifications of the Origin Space Telescope (OST) and its transit spectrograph \label{tab:ost}}
\tablehead{Items & Detector 1 & Detector 2 & Detector 3}
\startdata
\bf{OST} & \multicolumn{3}{c}{\bf{-}} \\
Primary mirror diameter (cm) & 924 & 924 & 924 \\
\hline
\bf{Transit spectrograph} & \multicolumn{3}{c}{\bf{-}} \\
Detector type & MCT & MCT & Si:As \\
Pixel format & $1,024 \times 1,024$ & $1,024 \times 1,024$ & $1,024 \times 1,024$ \\
Number of science pixels (pix) & 120,000 & 120,000 & 120,000 \\
Number of background pixels (pix) & 120,000 & 120,000 & 120,000 \\
Number of reference pixels (pix) & 760,000 & 760,000 & 760,000 \\
Radius of field of view (arcsec) & 2.0 & 2.0 & 2.0 \\
Transmittance inc. QE (\%) & 30 & 30 & 30 \\
Range of wavelength (\textmu m) & $3 \sim 6$ & $6 \sim 11$ & $11 \sim 22$ \\
Spectral resolution ($\lambda / \Delta \lambda$) & 100 & 100 & 100 \\
Exposure time at each frame (sec) & 60 & 60 & 60 \\
Standard deviation of gain fluctuation (ppm) & 100 & 100 & 100 \\
Dark current (electron/sec) & 1.0 & 1.0 & 1.0 \\
Readout noise (electron/read) & 5.5 & 5.5 & 5.5
\enddata
\end{deluxetable}

\subsection{Results} \label{subsec:results}

\begin{figure}[H]
\begin{center}
\includegraphics[width=12.5cm]{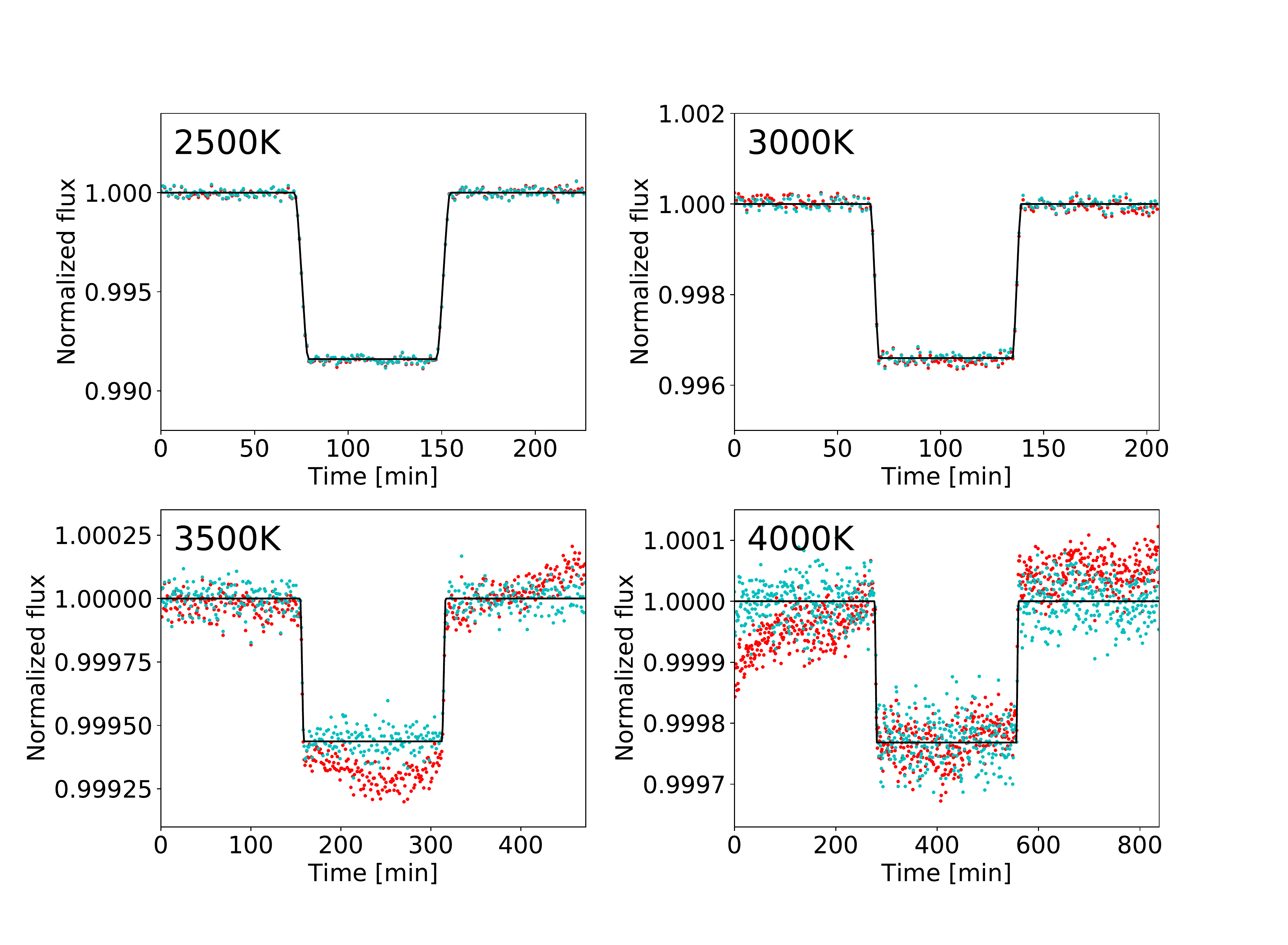}
\caption{Input transit light curves (black line), original transit data without calibration (red points) and calibrated transit data (cyan points) at the observing wavelength of 10 \textmu m for primary transits of transiting terrestrial planets orbiting four host stars with the different effective temperatures of 2,500, 3,000, 3,500, and 4,000 K, corresponding to the four simulation cases. \label{fig:transit_curve}}
\end{center}
\end{figure}

\begin{figure}[H]
\begin{center}
\includegraphics[width=12.5cm]{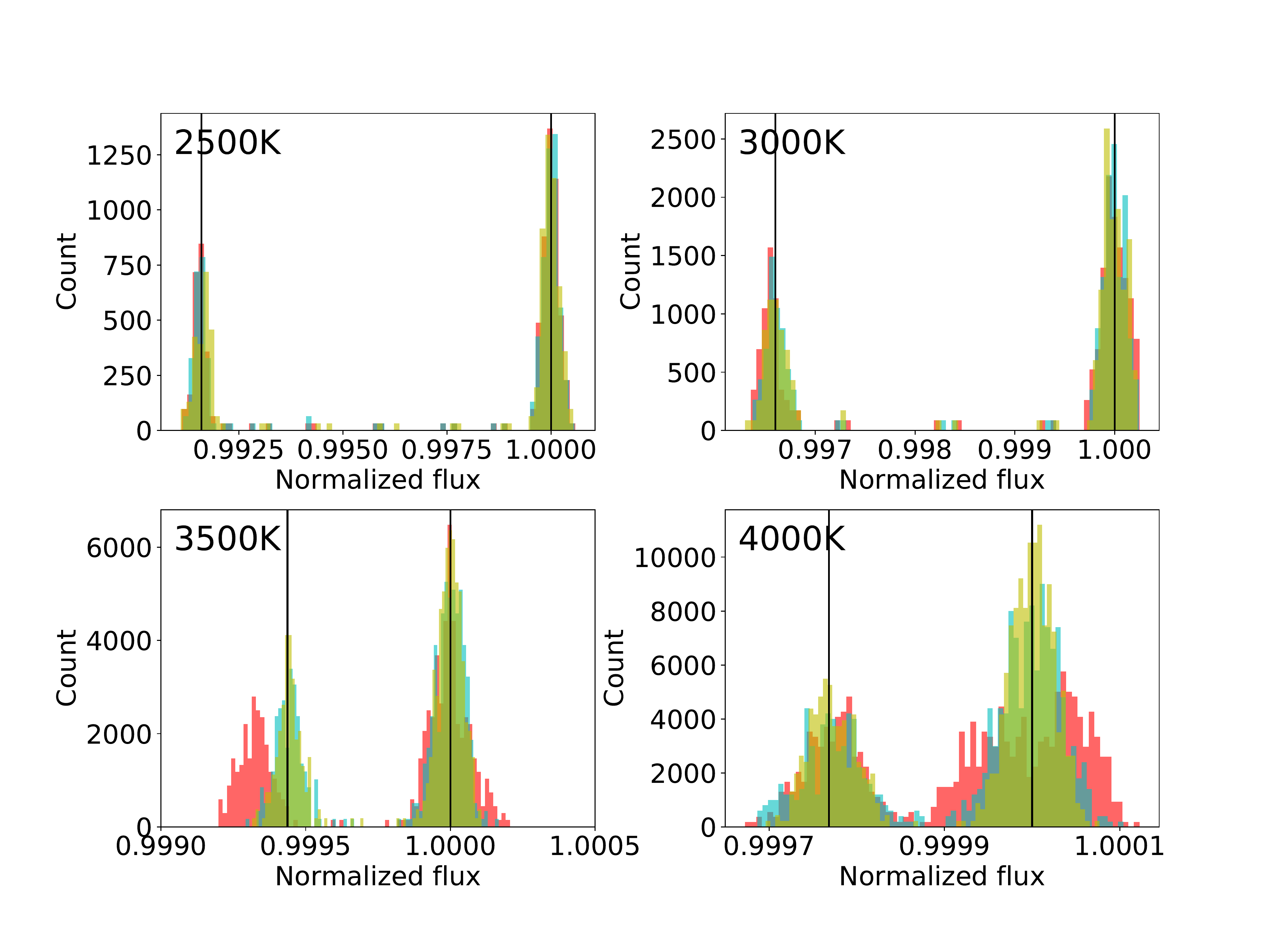}
\caption{Histograms of original transit data without calibration (red bins), calibrated data (cyan bins) and ideal transit data with only photon noise (yellow bins) at the observing wavelength of 10 \textmu m for the average transit data over 60 primary transits of the four simulation cases. The two vertical lines for each simulation case represent the two average values of the transit data in transit and out-of-transit, respectively. \label{fig:histogram_binned}}
\end{center}
\end{figure}

\begin{figure}[H]
\begin{center}
\includegraphics[angle=270, width=10cm]{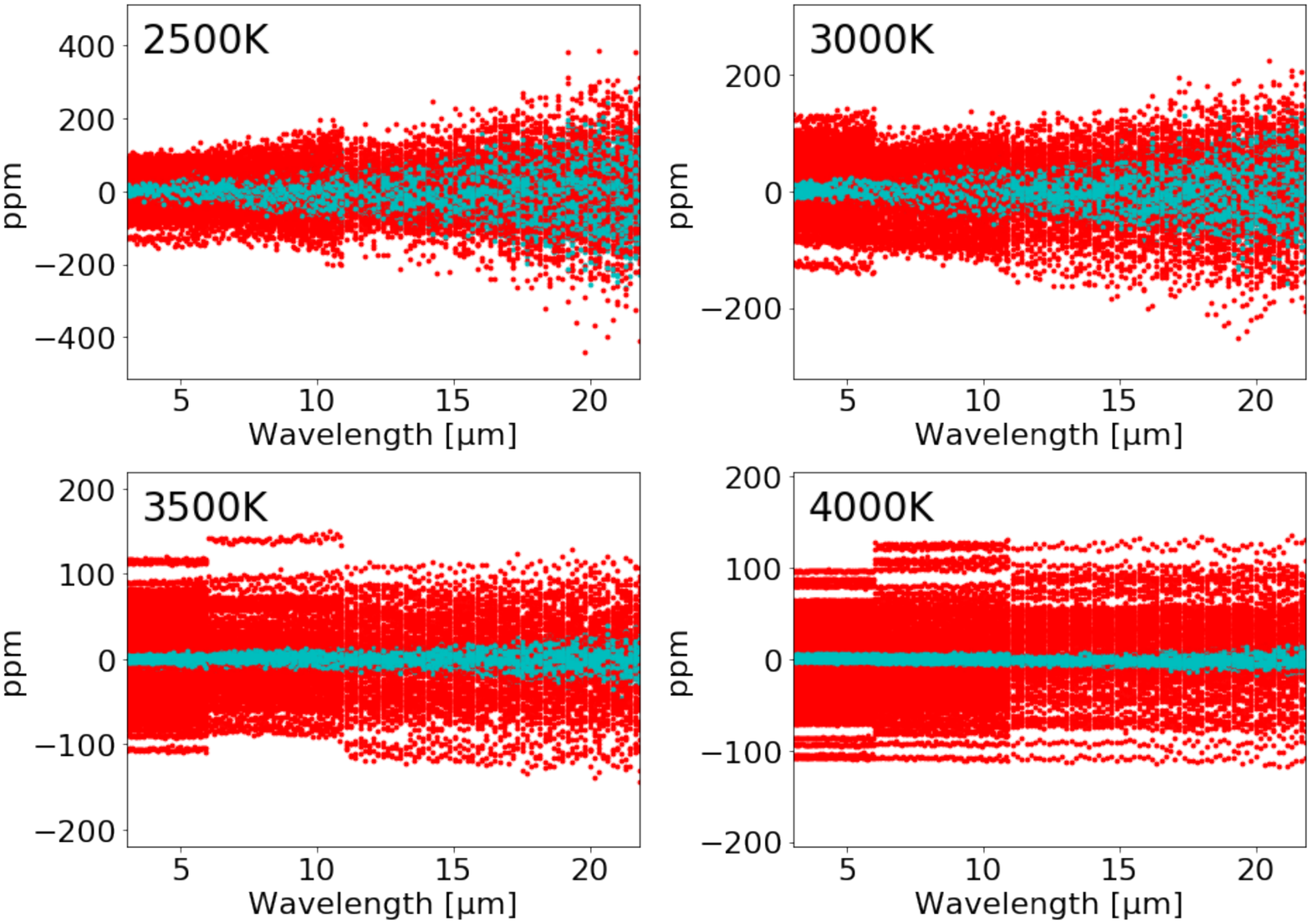}
\caption{Residuals between the input model data and the raw (red) or calibrated (cyan) one over the observing wavelength range of 3 to 22 \textmu m for the primary transits of the four simulation cases. The number of the iterations for each wavelength was 100. Because the mean of the plots for each wavelength equals to almost zero for the calibrated data, the systematic error associated to the raw data is reduced to the random component. The standard deviation of the plots for each wavelength is defined as the residual error. As the wavelength is longer, the residual error of the calibrated simulation system is larger because of the larger photon noise. The gaps seen in the early-M and late-K type stars in particular occur due to the limited number of the simulations. \label{fig:error}}
\end{center}
\end{figure}

\begin{figure}[H]
\begin{center}
\includegraphics[width=10cm]{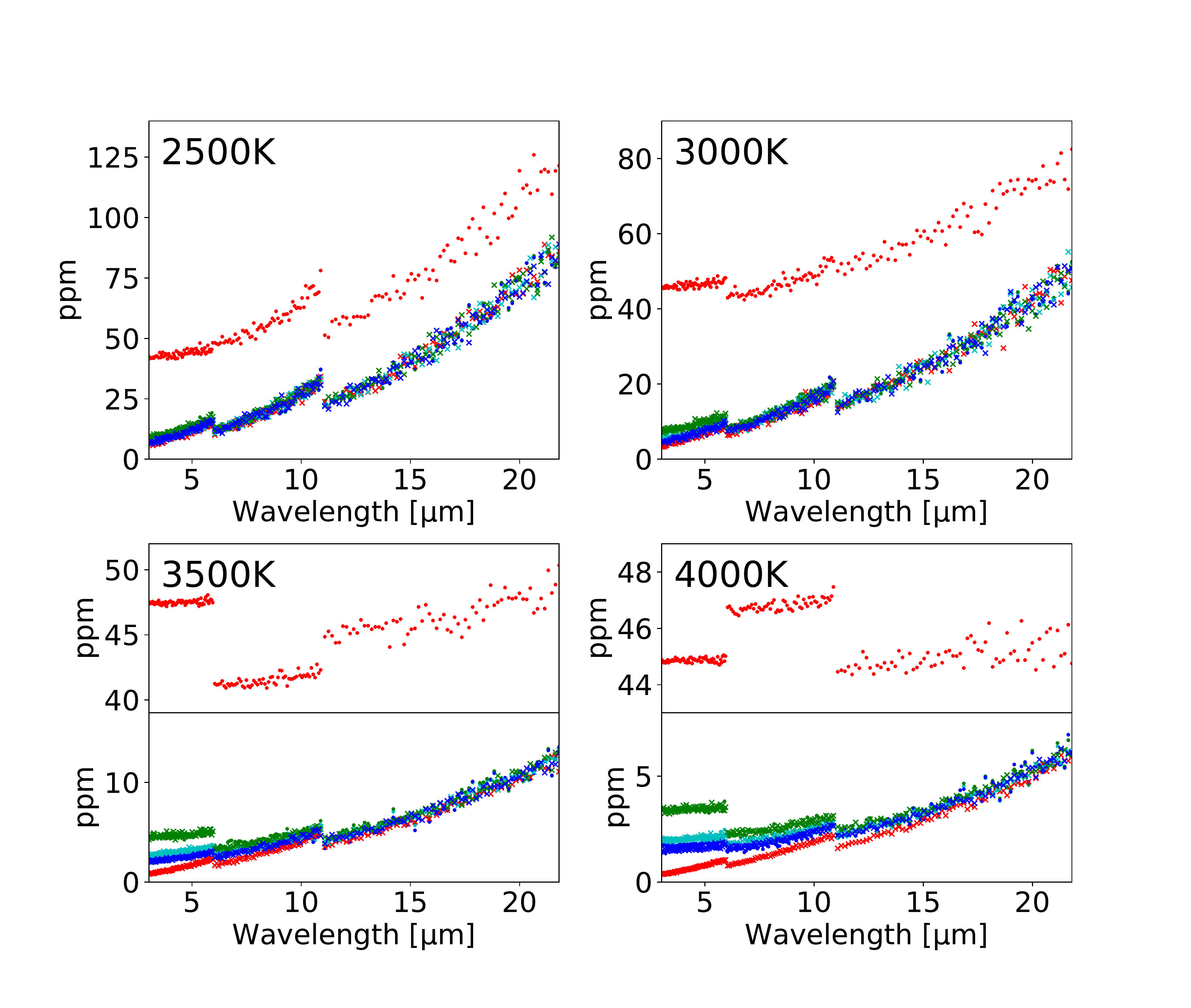}
\caption{Comparison of the raw data with three types of calibrated data over the wavelength range of 3 to 22 \textmu m for the primary transits in the four simulation cases. The red points represent the absolute values of the systematic errors attached to the raw data. The green, cyan, and blue points represent residual errors after calibration with three types of data: only background data, both of reference and background data, and only reference data, respectively. The red, green, cyan and blue crosses are the standard deviations of the random error before and after the calibration process with each signal, respectively. The residual error of the data reduced through the calibration is almost consistent with the random one for the four calibrated data because the systematic error associated to the raw data is reduced to the random component. \label{fig:evaluation_primary}}
\end{center}
\end{figure}

\begin{figure}[H]
\begin{center}
\includegraphics[width=10.5cm]{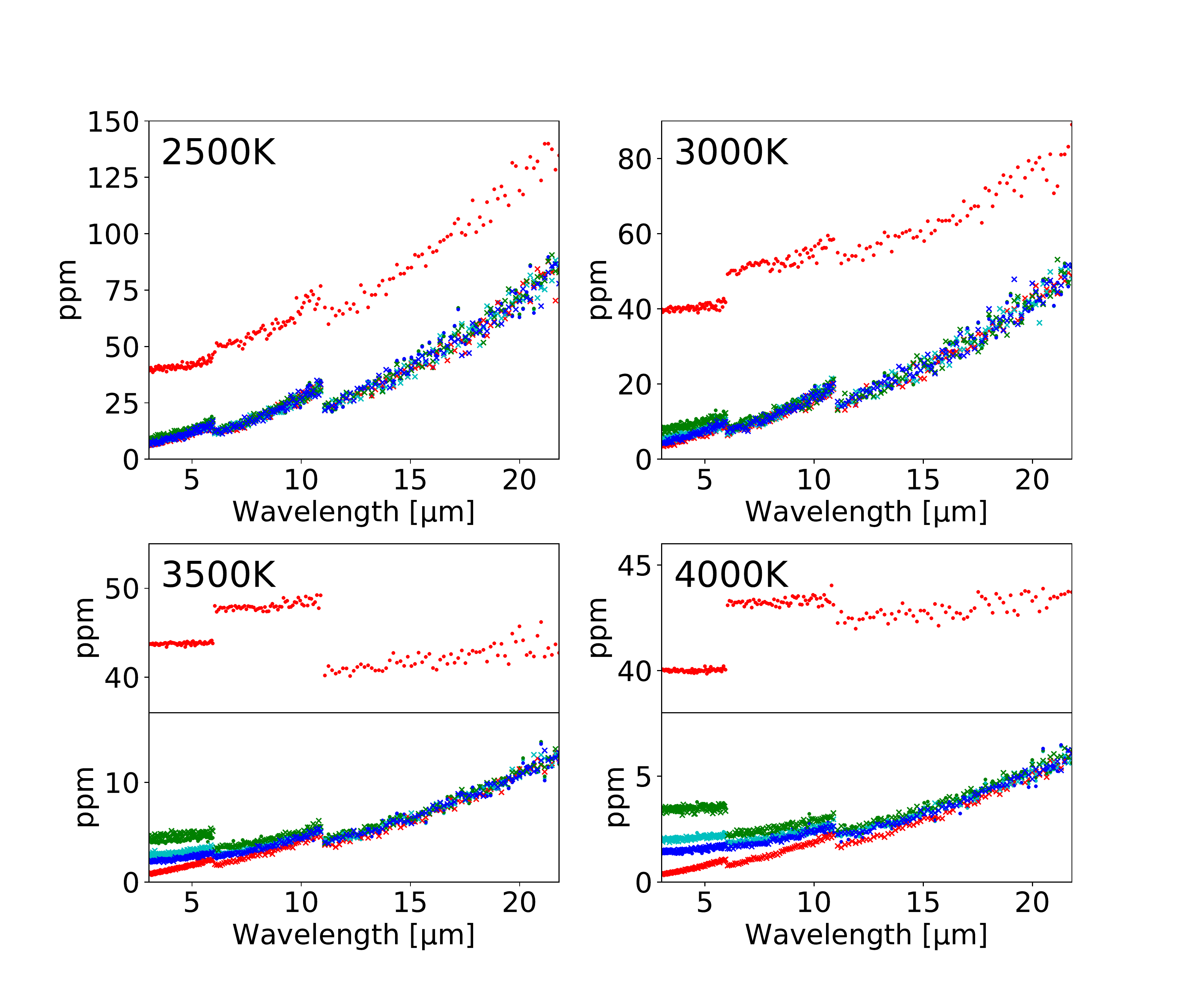}
\caption{Comparison of the raw data with three types of calibrated data over the wavelength range of 3 to 22 \textmu m for the secondary eclipses in the four simulation cases. The symbols are the same as those used for Figure \ref{fig:evaluation_primary}. Because the residual errors for the secondary eclipses are the same as those for the primary transits, the photometric accuracy after the calibration does not depend on the input signal. \label{fig:evaluation_secondary}}
\end{center}
\end{figure}

We performed numerical simulations to investigate how much the systematic components are reduced with the calibration technique under the assumptions described in Section \ref{subsec:setup}. The procedure for evaluation of the calibration method was as follows. First, a dataset composed of 60 transit light curves, enabling acquisition with a sufficient signal-to-noise ratio to characterize the atmospheres of transiting Earth-size planets orbiting M-type stars, was generated for each simulation case and the systematic components were reduced with the calibration method for each transit data. The 60 calibrated transit light curves were co-added for reduction of random noise. For example, the noise-to-signal ratio for the late-M type star was reduced to 25 ppm at 10 \textmu m by co-adding the 60 calibrated transit light curves. Figure \ref{fig:transit_curve} shows examples of the co-added transit light curves for the primary transits of transiting terrestrial planets orbiting the four host stars with different temperatures of 2,500, 3,000, 3,500, and 4,000 K. While the raw data for each simulation case (red points in Figure \ref{fig:transit_curve}) is largely distorted by the systematic component, the calibrated data (cyan points) is randomly distributed around the input transit curve (black solid line). Figure \ref{fig:histogram_binned} compares the histogram of the averaged calibration data at 10 \textmu m over the 60 transit light curves with those of the averaged original and ideal data that includes only shot noise for each simulation case. Although the calibration technique successfully removes the systematic components, the random error associated to the calibrated data is larger than that of the ideal data. This is because the calibrated data does not include only the shot noise acting on the original science pixels but also the random errors of the background and reference pixels. The systematic and random errors for the co-added data were evaluated. The systematic error for each dataset is defined as the residual between the depth of the model curve and that of the raw or calibrated data. The random error for each dataset is defined as the root-mean-square of the two standard deviations for the transit and out-of-transit. Last, the above procedure was iterated 100 times that enables to derive the systematic error of the proposed method with an accuracy better than the standard deviation of the photon noise. Figure \ref{fig:error} shows the residuals between the input model curve and the raw or calibrated data for the four simulation cases. The residual of the raw data along the wavelength at each iteration has an offset value and is randomly distributed around the offset. In contrast, the residuals for the calibrated datasets are randomly distributed around 0 and the systematic components are successfully reduced to the random ones. Therefore, the standard deviation of the residuals for each wavelength is defined as the residual error of the calibrated simulation system. Figures \ref{fig:evaluation_primary} and \ref{fig:evaluation_secondary} show the residual (systematic) and random errors in the wavelength range of 3 to 22 \textmu m at primary and secondary eclipses for each simulation case, respectively. The residual errors for both the primary and secondary eclipses are almost the same; the calibration accuracy does not depend on the type of the transit signal. In addition, the residual errors are almost consistent with the random ones of the raw data for the cases 1 to 3 of this simulation because the calibrated data is free from the systematic components in these cases. As the effective temperature of the host star increases, the residual error can be further reduced. This occurs because the shot noise increases for earlier spectral types of stars but the signal-to-noise ratio increases.

\begin{figure}[H]
\begin{center}
\includegraphics[width=10.5cm]{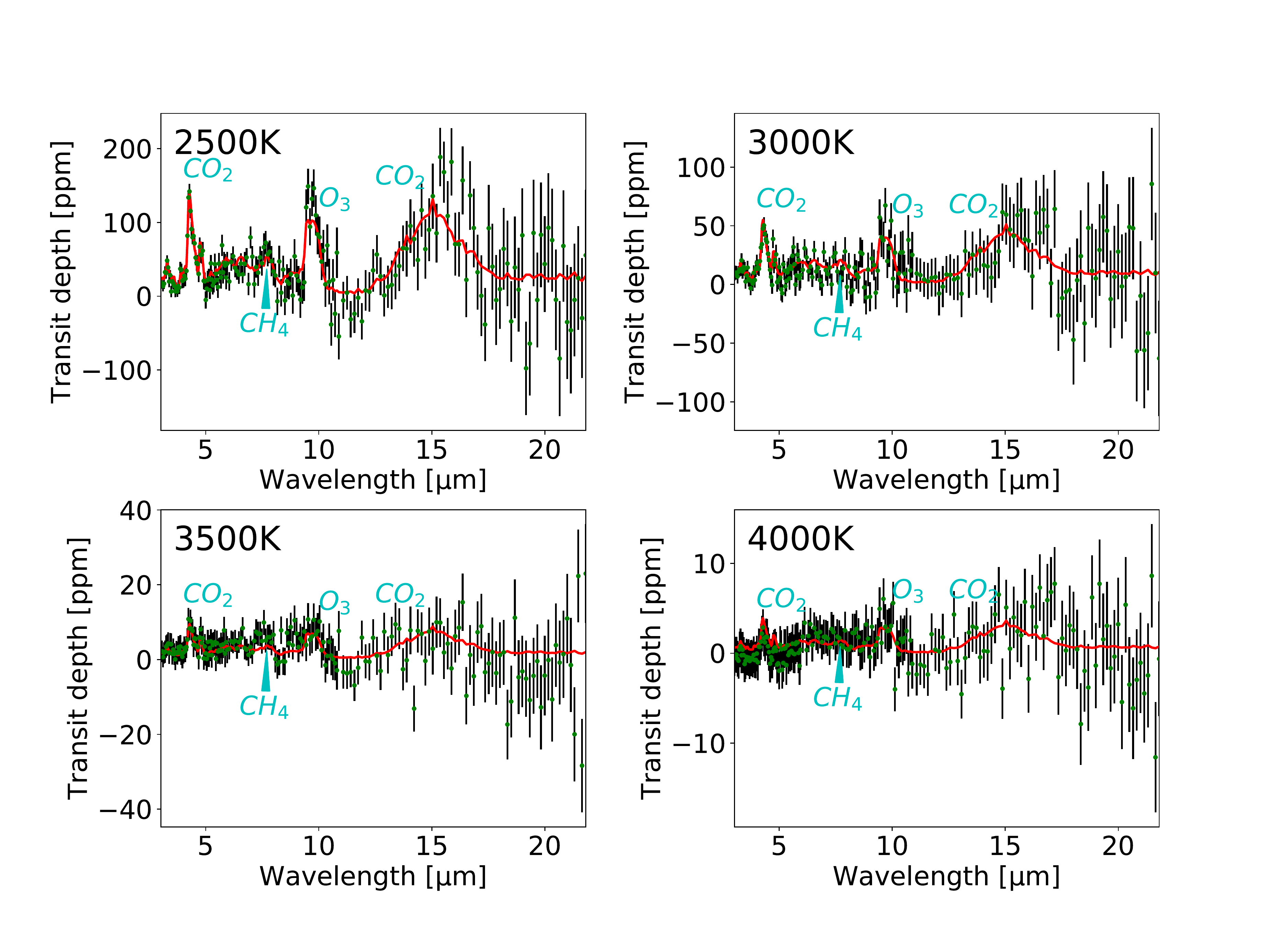}
\caption{Theoretical transmission spectra (red solid line) and the reconstructed spectra (green points) of Earth-like planets orbiting four host stars with the different effective temperatures of 2,500, 3,000, 3,500, and 4,000 K in the wavelength range of 3 to 22 \textmu m. The spectra are reconstructed through 60 transit observations with the densified pupil spectrograph mounted on a 9.3 m diameter telescope. The black vertical line attached to each green point represents the standard deviation of the random component associated to the calibrated data. The theoretical transmission spectra of the Earth-like planets were generated through the Planetary Spectrum Generator \citep{2015Sci...348..218V}. \label{fig:spectrum_primary}}
\end{center}
\end{figure}

\begin{figure}[H]
\begin{center}
\includegraphics[width=12cm]{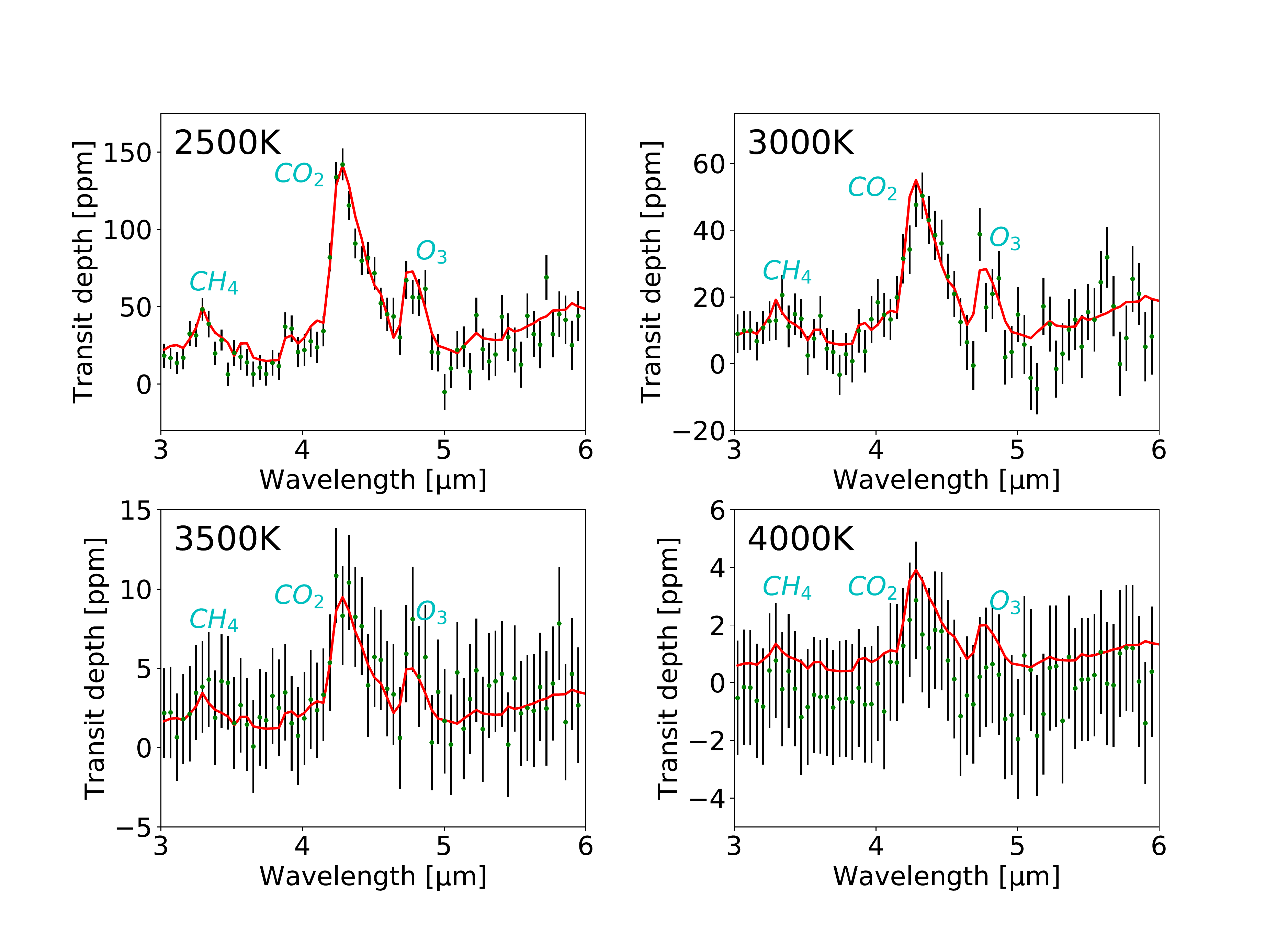}
\caption{Enlarged views of the short bands, 3 to 6 \textmu m, for the four simulation cases in Figure \ref{fig:spectrum_primary}. The symbols are same as those used for Figure \ref{fig:spectrum_primary}. \label{fig:spectrum_primary_short}}
\end{center}
\end{figure}

\begin{figure}[H]
\begin{center}
\includegraphics[width=12cm]{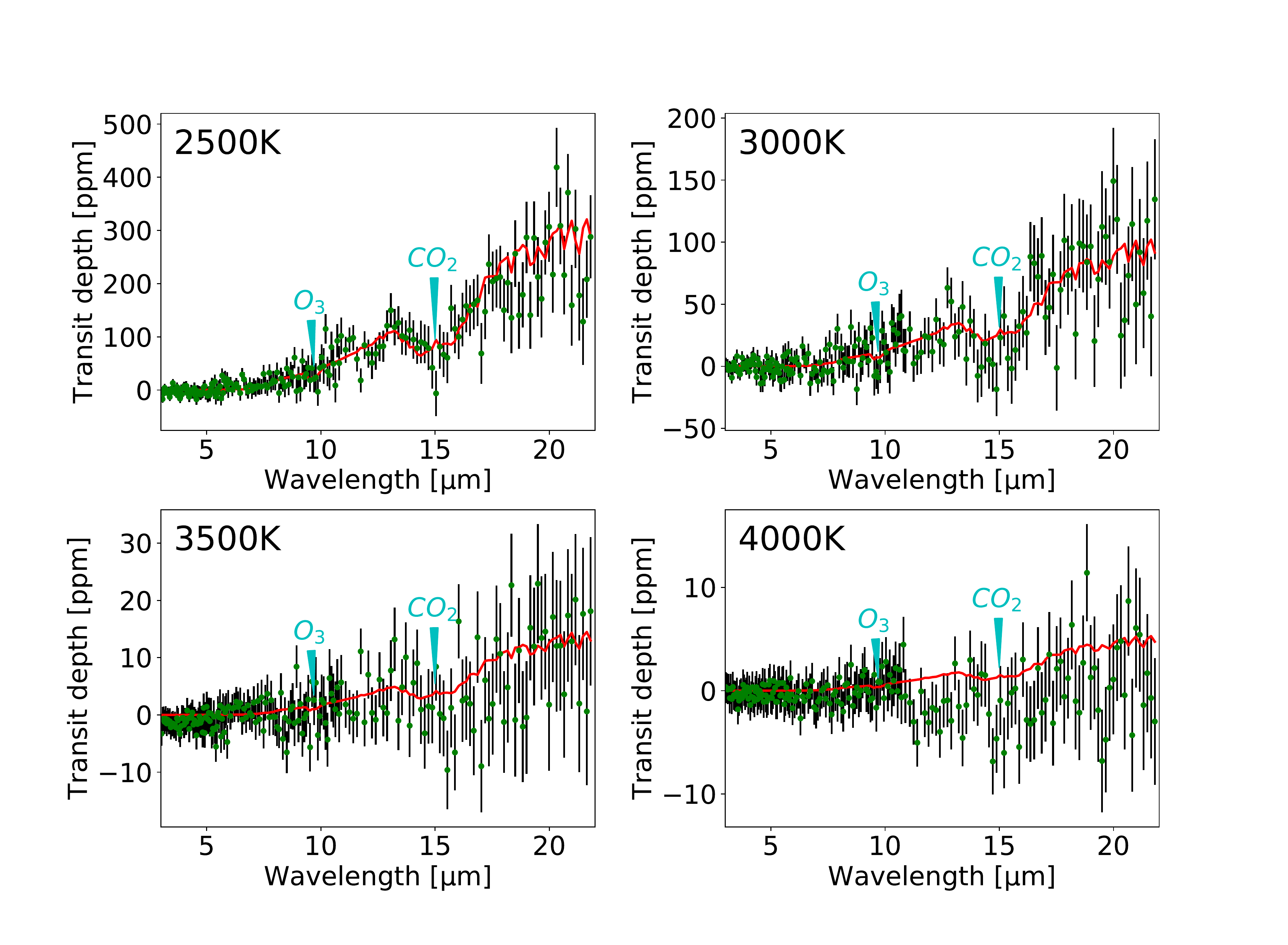}
\caption{Theoretical emission spectra (red solid line) and reconstructed spectra (green points) of Earth-like planets orbiting four host stars with the different effective temperatures of 2,500, 3,000, 3,500, and 4,000 K in the wavelength range of 3 to 22 \textmu m. The symbols are the same as those used for Figure \ref{fig:spectrum_primary}. \label{fig:spectrum_secondary}}
\end{center}
\end{figure}

Once the residual (systematic) and random errors of this simulation system were evaluated, the spectrum data including the derived errors was compared with the theoretical transmission and emission spectra of a transiting Earth-like planet for each simulation case. The atmospheric composition of the transiting planet was similar as one of the Earth. There are a number of absorption lines over 3 to 22 \textmu m: $\rm CH_4$ at 3.3 and 7.6 \textmu m, $\rm O_3$ at 4.75 and 9.6 \textmu m, $\rm CO_2$ at 4.3 and 15 \textmu m and $\rm H_{2}O$ around 6.3 and beyond 17 \textmu m, respectively. The transmission and emission spectra of the four target planets were generated through the planetary spectrum generator \citep{2015Sci...348..218V}. Figures \ref{fig:spectrum_primary} compares the theoretical transmission spectra of the Earth-like planet with the reconstructed one for the four simulation cases over wavelength ranges of 3 to 22 \textmu m; Figure \ref{fig:spectrum_primary_short} shows an enlarged view of the short band, 3 to 6 \textmu m, in Figure \ref{fig:spectrum_primary}. In addition, Figure \ref{fig:spectrum_secondary} compares of the emission spectra of the Earth-like planet with the reconstructed spectra for the four simulation cases. The important absorption features including $\rm H_{2}O$, $\rm CH_{4}$, $\rm O_{3}$ and $\rm CO_{2}$ can be detected from the reconstructed transmission spectra of planets orbiting the M-type stars. Measurement of both oxidized species $\rm O_3$ and reduced species of $\rm CH_4$ may confirm whether non-equilibrium atmosphere indicates biological activity on a planet. In addition, the effective temperatures of the planets, as information concerning habitability, can be also measured through secondary eclipses. In contrast, the absorption features are embedded in the systematic and random components for the late-K type star. Thus, this proposed calibration method contributes to the measurement of the habitability and biosignature of the nearby Earth-like planets orbiting late-type stars through both the transmission spectroscopy and secondary eclipse of the planets.

\section{Discussion} \label{sec:discussion}

\subsection{Dependence of the signal-to-noise ratios of background and reference pixels upon calibration accuracy} \label{subsec:dependency}

In Section \ref{sec:simulation}, we used the background and reference data to calibrate the time-variation components. The calibration accuracy depends upon the signal-to-noise ratios of the background and reference pixels. Note that, while the signal-to-noise ratio of each background pixel changes with the radius of the field of view as well as the brightness of the zodiacal light, each reference pixel depends upon the level of the dark current and readout noise. In this subsection, we discuss how the spectro-photometric accuracy changes for various signal-to-noise ratios of the background and reference pixels, respectively.

\subsubsection{Signal-to-noise ratio of the background pixels} \label{subsubsec:sn_zodi}

While the proposed calibration method reduces the systematic noise to random noise for the late- and middle-M type stars, slight offsets remain for the early-M and late-K type stars. This is because the second-order fluctuation represented in Equation (\ref{equ:v_sub,norm}) still remains after the calibration and contributes to the calibrated data as systematic noise. Here, the second-order fluctuation is expected to be minimized when the calibrated data become closer to the ideal photon-noise-limit thanks to a high signal-to-noise ratio of the background pixel. In contrast, as the signal-to-noise ratio of the background pixel increases (i.e., the number of the incident photons increases in the background pixel), the photon noise from the background light  increases in the science pixel; the signal-to-noise ratio of the science pixel decreases.

\begin{figure}[H]
\begin{center}
\includegraphics[width=12cm]{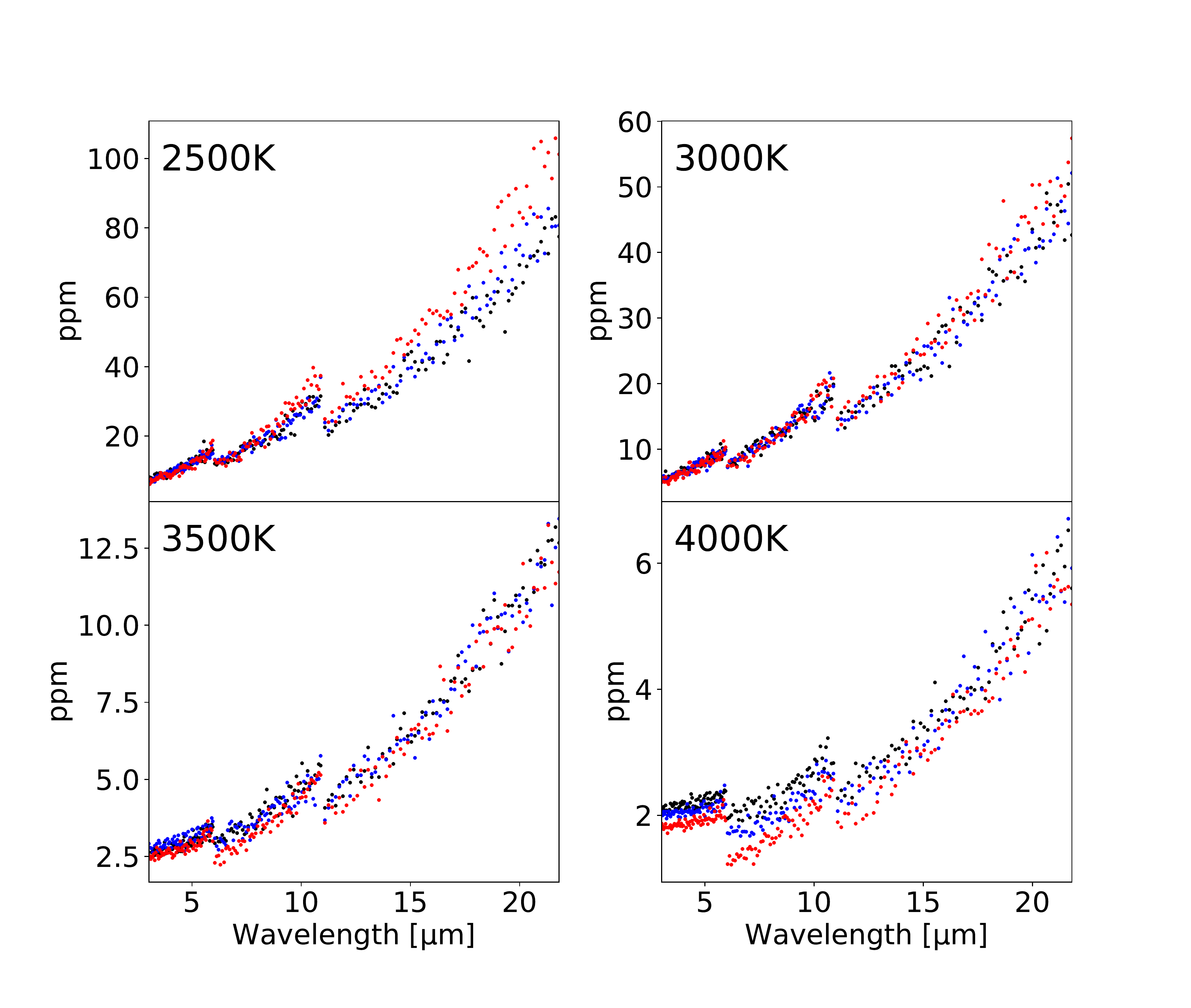}
\caption{Residual systematic noises of the calibrated data with both the background and reference pixels over the wavelength range of 3 to 22 \textmu m for the primary transits of the four simulation cases. Fields of view are 1.0 (black points), 2.0 (blue points), and 4.0 arcsec (red points) in radius. Because the residual errors associated with the calibrated data include the shot noise of the zodiacal light, the residual error is further reduced as the radius of the field of view increases, except over the longer wavelength range for late- and middle-M type stars. \label{fig:sn_zodi}}
\end{center}
\end{figure}

We have examined how the level of the background light or the radius of the field stop affects the calibration accuracy. Figure \ref{fig:sn_zodi} shows how the photometric accuracy for each spectral element depends upon the radius of the field of view. Note that the larger field of view is equal to an increase in the zodiacal light. For the case of the larger field of view, while the photometric accuracy for each spectral element can be improved over the shorter wavelength range thanks to the higher signal-to-noise ratio of the background pixel, the photometric accuracy worsens in the longer wavelength range, especially for the late- and middle-M type stars, because of the larger photon noise in the science pixel.

\subsubsection{Signal-to-noise ratio of the reference pixels} \label{subsubsec:sn_dark}

The signal-to-noise ratio of the reference pixel depends upon the level of the dark current and the readout noise. The calibration accuracy is affected by the signal-to-noise ratios of the reference and background pixels. Although the dark current was fixed to 1.0 e-/sec in the previous simulation based on the performance of the MCT and Si:As detector systems, the dependency of the dark current level on the calibration method has been investigated.

\begin{figure}[H]
\begin{center}
\includegraphics[width=12cm]{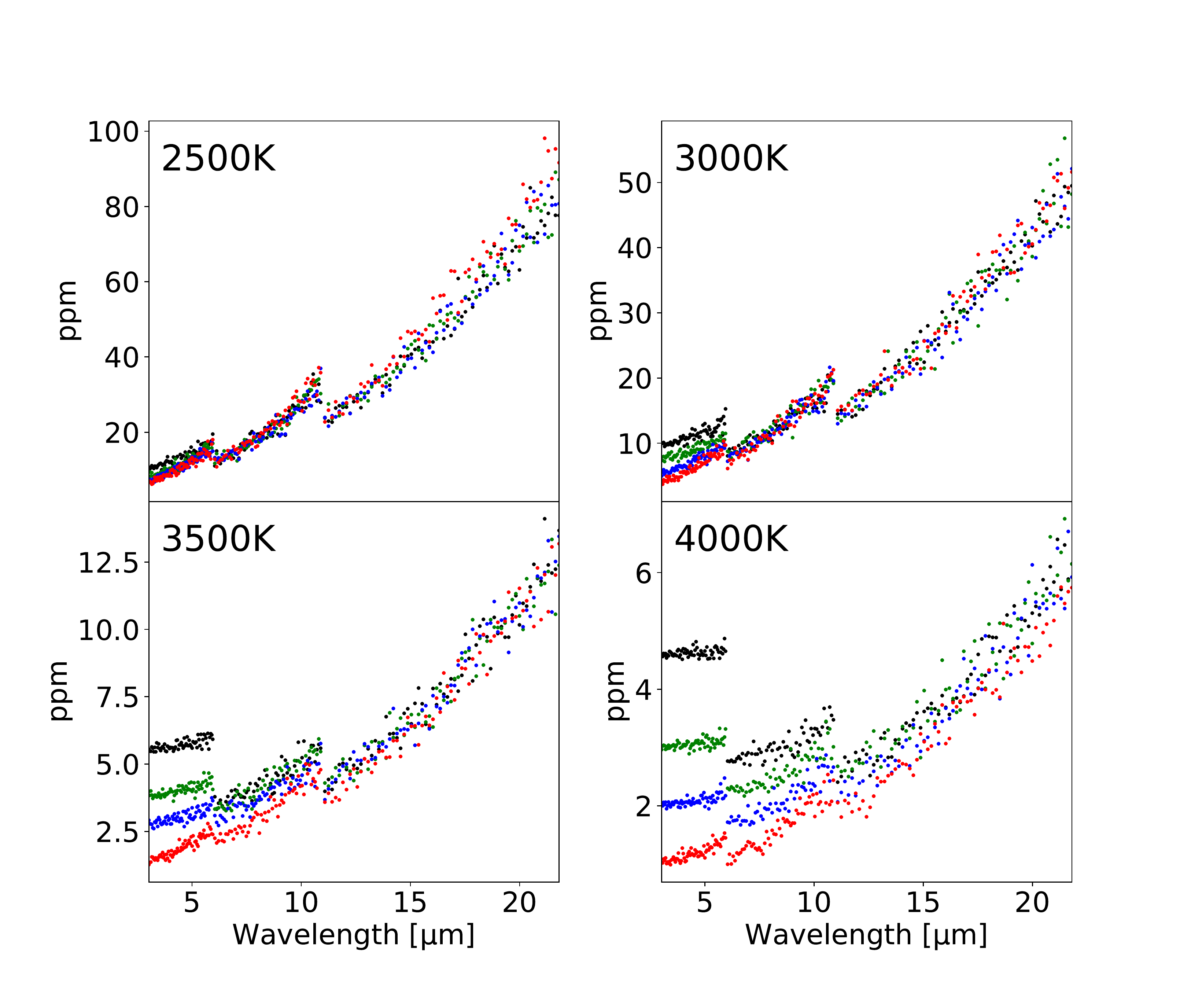}
\caption{Residual systematic noises of the calibrated data over the wavelength range of 3 to 22 \textmu m in the primary transits of each simulation case for the various dark current levels of 0.2 (black points), 0.5 (green points), 1.0 (blue points), and 5.0 e-/s (red points). Because the residual errors associated to the calibrated data include the shot noise of the dark current, the residual error is further reduced as the dark current increases. \label{fig:sn_dark}}
\end{center}
\end{figure}

Figure \ref{fig:sn_dark} shows the residual errors associated to the calibrated data for various levels of the dark current for the primary transit of each simulation case with various dark currents of 0.2, 0.5, 1.0, and 5.0 e-/sec. The higher dark-current level improves the calibration technique except over the longer wavelength range of late- and middle-M type stars. Thus, the dark current level is an important parameter for transit spectroscopy around the early type of stars. Note that, because higher dark current level requires higher temperature of the detector system \citep[e.g.,][]{2003SPIE.4850..890E}, the higher dark current may have an impact on cryogenic space observatories such as the Origins Space Telescope.

\subsection{Impact of time-variation of background light and dark current upon the calibration method} \label{subsec:time_variation}

So far, we have assumed that the zodiacal light and the dark current are constant; however, actually, the signals have small time-variation components. Here, we discuss how these time-variations affect the proposed method.

\subsubsection{Time-variation of background light} \label{subsubsec:variation_zodi}

In previous sections, the background light was treated as a constant value; the background component, except for the zodiacal and galactic planes, is thought to be smooth on the timescale of a transit observation because the spatial distribution of zodiacal dust is mainly determined by the gravity field of the Sun and the Solar-system planets \citep{2016AJ....151...71K}. However, since the spatial variation of the zodiacal light at a high spatial scale ($\sim$1") is as yet unrevealed due to a lack of high-angular resolution observations at mid-infrared wavelengths, there may be time-variations of the zodiacal light over a timescale of a few hours due to its spatial brightness variation.

\begin{figure}[H]
\begin{center}
\includegraphics[width=12cm]{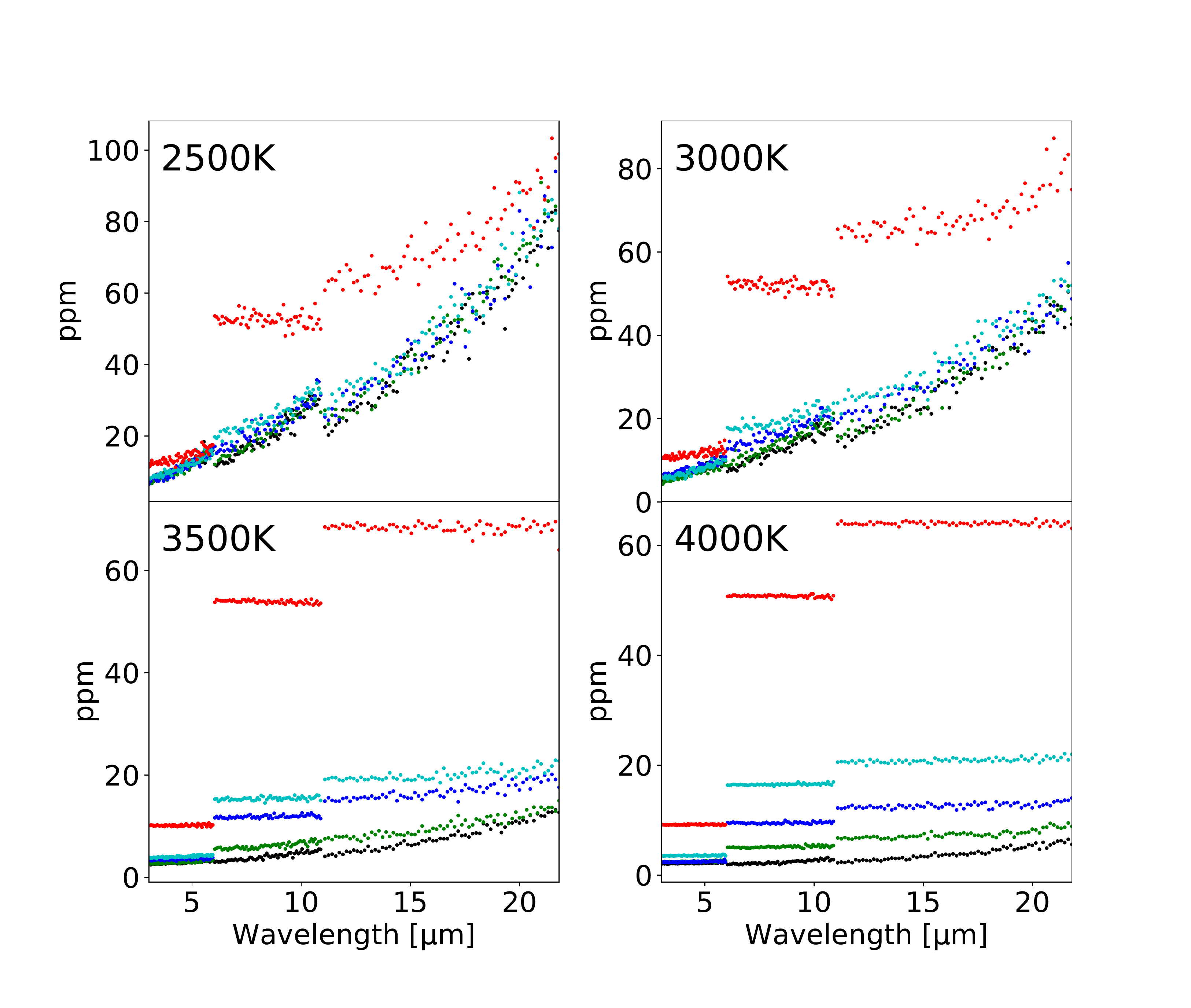}
\caption{Residual systematic noises of the calibrated data with both background and reference pixels over the wavelength range of 3 to 22 \textmu m for the primary transits of the four simulation cases with various standard deviations of the time-fluctuation of zodiacal light. The black, green, blue, cyan, and red points show the residual systematic errors for the standard deviations of 0, 100, 200, 300, and 1,000 ppm, respectively. \label{fig:variation_zodi}}
\end{center}
\end{figure}

Here, we considered the effect of the small time-variation of the background pixels upon the proposed method and performed numerical simulations to evaluate how the effect degrades the calibration accuracy. Figure \ref{fig:variation_zodi} shows the residual systematic errors associated to the calibrated data in the primary transit of each simulation case for various standard deviations of the zodiacal light time-variation of 0, 100, 200, 300, and 1,000 ppm. The time-variation of a few hundred ppm can be ignored for the late- and middle-M type stars. By contrast, for the early-M and late-K type stars, the residual systematic errors are larger in the longer wavelength range because larger time-variation occurs due to the brighter zodiacal light.

\subsubsection{Time-variation of the reference pixels} \label{subsubsec:variation_dark}

\begin{figure}[H]
\begin{center}
\includegraphics[width=12cm]{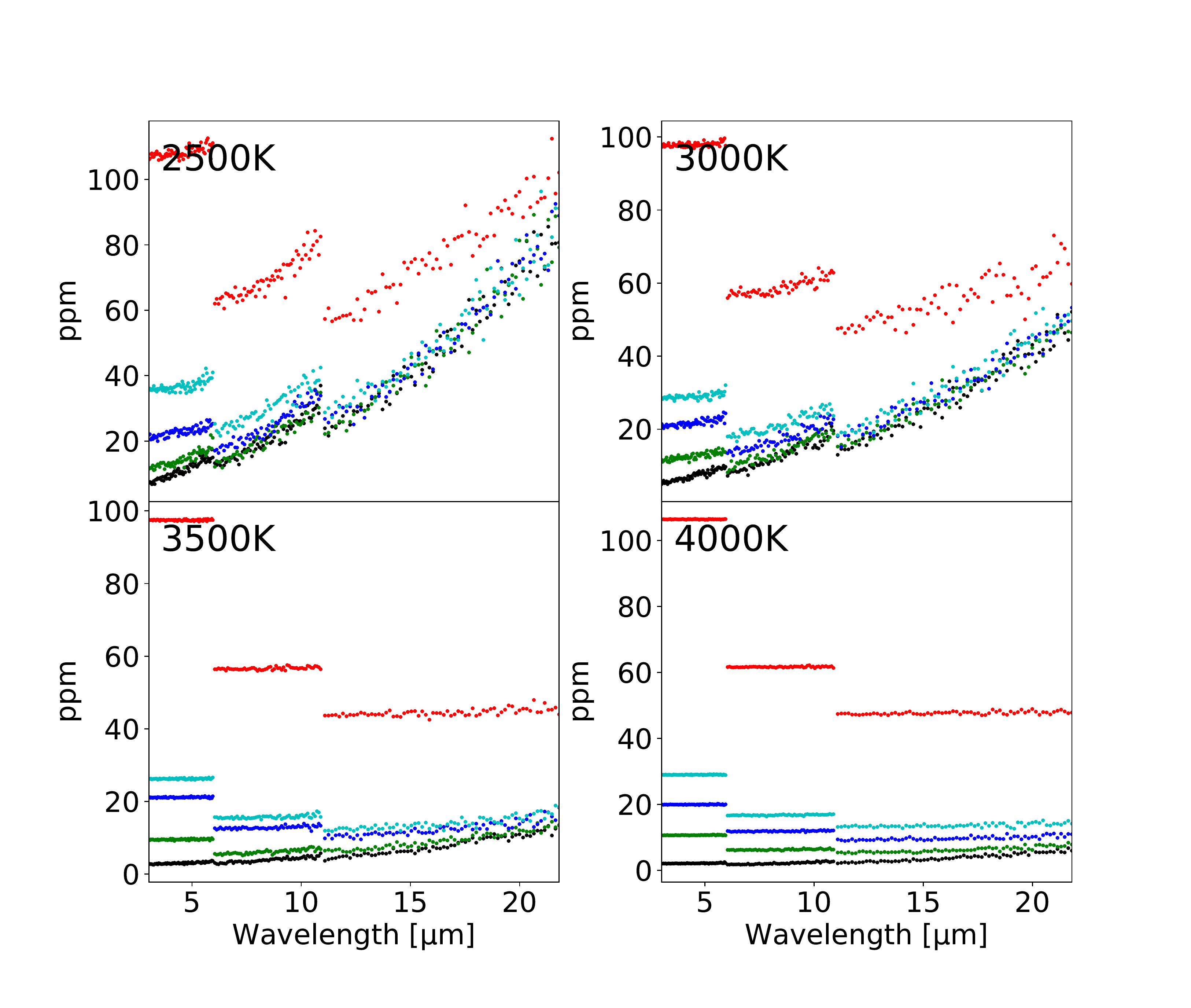}
\caption{Residual systematic noises of the calibrated data over the wavelength range from 3 to 22 \textmu m in the primary transits of each simulation case for various standard deviations of the time-fluctuation of the dark current. The black, green, blue, cyan, and red points show the residual systematic errors for standard deviations of 0, 100, 200, 300, and 1,000 ppm, respectively. \label{fig:variation_dark}}
\end{center}
\end{figure}

Since the dark current in the ROIC depends on the detector temperature, we should consider the impact of time-variation of the dark current upon the proposed method due to small variation of the detector temperature. We evaluated how the degradation of the calibration accuracy due to time-variation of the dark current. Figure \ref{fig:variation_dark} shows the residual systematic errors associated with the calibrated data in the primary transit of each simulation case for standard deviations of the dark current time-variation of 0, 100, 200, 300, and 1,000 ppm. The effect of time-variation of the dark current cannot be ignored when the standard deviation is more than 100 ppm, because the signal-to-noise ratio of the averaged reference data over all reference pixels for each exposure is around 100 ppm. In addition, the time-variation of the dark current largely affects the shorter band because the ratio of reference to science data that is used for reconstruction of the common time-variation from the reference data is larger in the shorter wavelength range (see Equation (\ref{equ:v_cal})).

There are two approaches for minimizing the impact of the dark current time-variation upon the method. The first is to stabilize the detector temperature and the second is to measure the detector temperature and calibrate the dark current. Because the former approach is technically more difficult than the latter, the latter is discussed in this paper. Based on previous studies \citep[i.e.,][]{2003SPIE.4850..890E, 2013SPIE.1306..6978}, the relationship between the dark current and the detector temperature (i.e., the Arrhenius plot) can be expressed as below:
\begin{equation}
I_{dark} = C_{1}10^{\frac{C_2}{T}} ,
\end{equation}
where $C_1$ and $C_2$ are constant values that depend upon the type of detectors and the detector temperature. When the dark current is 1 e-/s, the detector temperature should be measured down to 0.15 mK at 30 K for the MCT detector and 0.02 mK at 8 K for the Si:As detector to determine the dark current time-variation with an accuracy of 100 ppm. Here, focusing on the fact that the proposed method does not require the absolute accuracy of the detector temperature but the relative one, commercially available cryogenic temperature sensors \citep[i.e.,][]{2001IEEE...1...52} can meet the above strict requirements on the relative-temperature-measurement mode \citep{1994OMEGA}. The resolvable temperature in the relative-temperature-measurement mode can be enhanced by a factor of $\frac{V}{\Delta V}$ compared to that of the absolute mode, where $V$ is the voltage required for a full measurement range in the absolute-temperature-measurement mode and $\Delta V$ is the variation of voltage during measurement. Thus, two cryogenic temperature sensors with absolute- and relative-temperature-measurements may be sufficient to establish the proposed method.

\subsection{Three unconsidered factors} \label{subsec:three_factors}

There are four factors that are not considered for the numerical simulations discussed in the previous section. The first unconsidered factor is the estimation error of the offset components that is caused by adding residual carriers present in the last frame to those in the current frame. These residual carriers are generated by the electron capturing in the blocking layer of the semiconductor device and dielectric relaxation because of its long RC time constant \citep{2015PASP..127..665R} and the persistence effect in the multiplexers of the ROIC \citep{2014PASP..126.1134B}. The latent effect also leads to the estimation error of the offset voltage, thus affecting the calibration method. The second factor is second-order effect of the metal-oxide-semiconductor field effect transistors (MOSFETs) in the ROIC. The second-order effect mainly arises from the following four phenomena: channel length modulation effect, substrate effect, short channel effect, and sub-threshold region effect. The second-order effects also generate artifact time-variation components. The last systematic error is observed as a steep ramp in the beginning of the observations. The steep ramp arises from a gradual decrease in the recapture rate of photoelectrons in the semiconductor device. As introduced in Section \ref{sec:concept}, because the recapture rate depends on the number of holes that capture the photoelectrons, the recapture rate decreases as the photoelectrons bury the holes. As the target object is brighter and the observing wavelength is closer to the peak of the spectrum, the recapture rate is thought to more rapidly decrease and further affect the measurement of the transit spectroscopy. Note that, although the decrease of the photometry curve over a timescale of ten hours after the ramp may also affect phase-curve measurements, the impact of the fallback on transit spectroscopy is not discussed here because its amplitude is smaller than that of the ramp.

\begin{figure}[H]
\begin{center}
\includegraphics[width=12cm]{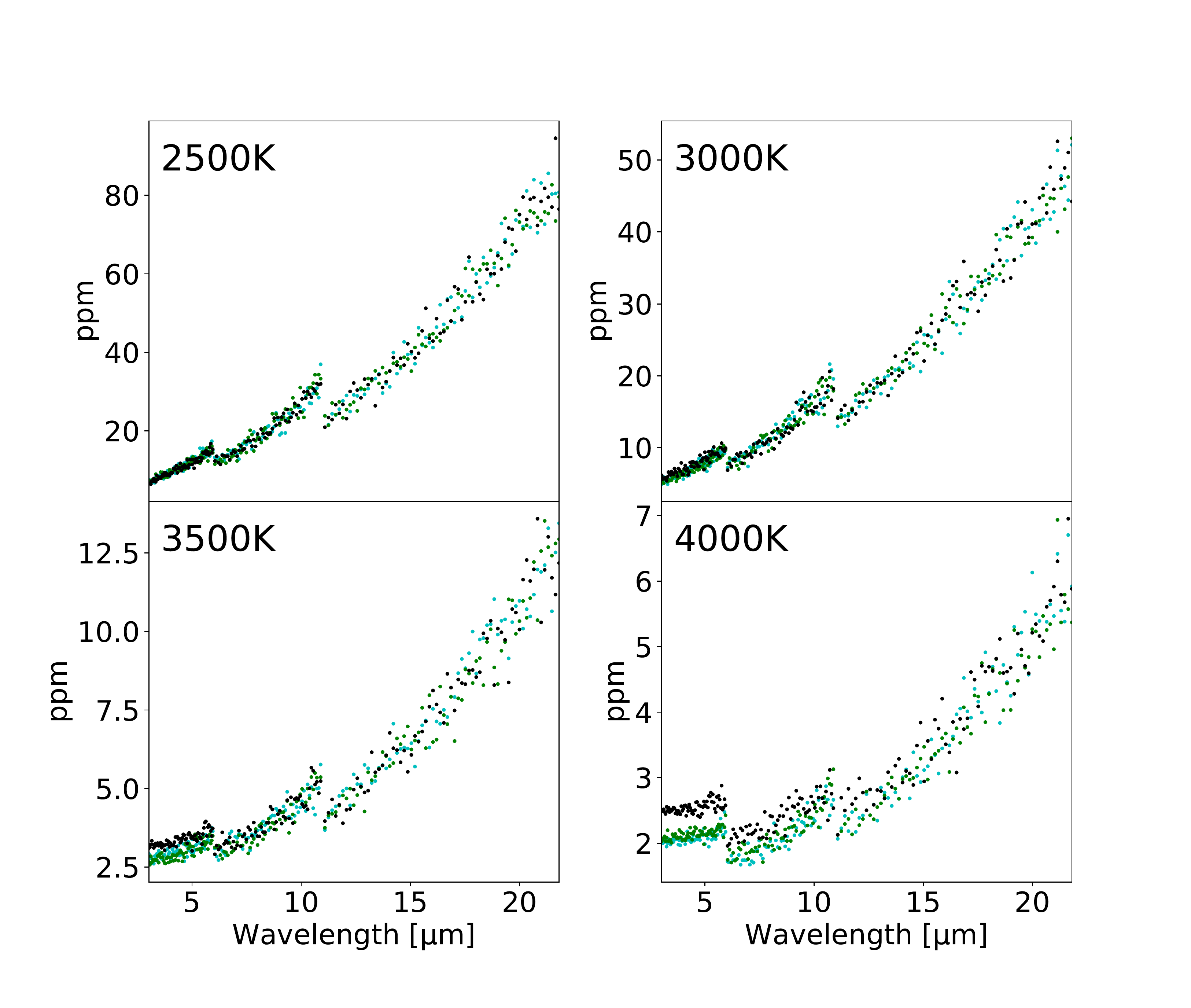}
\caption{Residual systematic noises of the calibrated data over the wavelength range of 3 to 22 \textmu m in the primary transits of each simulation case for estimation errors of the offset voltage of 0 (cyan points), 10 (green points), and 50 \% (black points). \label{fig:offset_error}}
\end{center}
\end{figure}

As discussed in Section \ref{sec:concept}, densified pupil spectroscopy that forms multiple spectra on the detector plane can mitigate various systematic errors. However, the systematic components potentially reduce the usefulness of the calibration method. The above three components are categorized into two. One is the latent effect and second-order effect that contribute to the estimation error of the offset voltage. The other is a gradual increase in the effective gain due to electron trapping at the beginning of the transit observation. When the latent and second-order effects produce the offset voltages, ${\Delta}V_{back}$ and ${\Delta}V_{ref}$, in the background and reference pixels, respectively, Equation (\ref{equ:v_sub,norm}) without the quadratic term changes to
\begin{equation}
v_{sub,norm}(\lambda,t) \approx 1+\frac{{\delta}f_{s}(\lambda,t)}{\hoge<f_{s}(\lambda)>_{t}}+\sigma_{sub,norm}(\lambda)+v_{offset}(\lambda,t) ,
\end{equation}
where $v_{offset}$ is an offset component in the transit curve normalized by the source flux:
\begin{equation}
\label{equ:v_offset}
v_{offset}(\lambda,t) \approx \left(1-\frac{\hoge<A_{com}(t)>_{t}\{f_{z}(\lambda)+2\hoge<I_{dark,ij}>_{pix}\}}{\hoge<A_{com}(t)>_{t}\{f_{z}(\lambda)+2\hoge<I_{dark,ij}>_{pix}\}+{\Delta}V_{back}+{\Delta}V_{ref}}\right)\left\{1+\frac{f_{z}(\lambda)+\hoge<I_{dark,ij}>_{pix}}{\hoge<f_{s}(\lambda,t)>_{t}}\right\}\frac{{\delta}A_{com}(t)}{\hoge<A_{com}(t)>_{t}} .
\end{equation}
The estimation error of the offset voltage for the pixel-averaged background and reference pixels leaves the systematic component of $\frac{{\delta}A_{com}(t)}{\hoge<A_{com}(t)>_{t}}$ in the calibrated transit data. As shown in Equation (\ref{equ:v_offset}), when the gain variation, $\frac{{\delta}A_{com}(t)}{\hoge<A_{com}(t)>_{t}}$, is small, this estimation error can be ignored. Figure \ref{fig:offset_error} shows the residual systematic errors over the wavelength range of 3 to 22 \textmu m for the various estimation errors of the offset voltage. The effect of the estimation error of the offset voltage on the proposed method can be ignored except for the short wavelength ranges of the early-M and late-K type stars. 

Last, we discuss the impact of the increase of the effective gain arising from the decrease in the recapture rate of photoelectrons on the transit spectroscopy. According to a previous study \citep{2012ApJ...752...81C}, the time-variation of the signal due to the electron trapping can be formulated as a function of time, $\beta(t)$:
\begin{equation}
\beta(t) = (1-ae^{-\frac{t}{\tau_{1}}}){\times}e^{-\frac{t}{\tau_{2}}} ,
\end{equation}
where $\tau_{1}$ and $\tau_{2}$ are timescales of the ramp and the fallback with $\tau_{2}>\tau_{1}$, respectively, and $a$ is a free parameter depending on the signal value. Here, focusing on the fact that the number of photons falling on each pixel from a target object is much reduced owing to the multiple spectra formed on the detector plane, the timescales of the ramp and fallback are much longer than that of the general-purpose spectrograph:
\begin{equation}
\label{equ:beta}
\beta(t) \approx 1-ae^{-\frac{t}{\tau_{1}}} .
\end{equation}
$\tau_{1}$ for the densified pupil spectrograph is re-calculated using $\tau_{1}$ and $a$ derived for the transit observations of the HD 209458 system conducted with the Multi-band Imaging photometer for Spitzer \citep{2012ApJ...752...81C} (Table \ref{tab:beta}). Note that $a$ is fixed and $\tau_{1}$ is assumed to be inversely proportional to the number of photons falling on a pixel. When $\beta_{sci}(t)$ in the science pixel has a fluctuation component with an amplitude of ${\delta}\beta_{sci}(\lambda,t)$ over one transit observation, Equation (\ref{equ:v_sub,norm}) without the quadratic term is rewritten as
\begin{equation}
v_{sub,norm}(\lambda,t) \approx 1+\frac{{\delta}f_{s}(\lambda,t)}{\hoge<f_{s}(\lambda)>_{t}}+\sigma_{sub,norm}(\lambda)+v_{beta}(\lambda,t) ,
\end{equation}
where $v_{beta}(\lambda,t)$ is the residual error caused by the fluctuation of $\beta_{sci}(t)$:
\begin{equation}
v_{beta}(\lambda,t) \approx \left\{1+\frac{f_{z}(\lambda)+\hoge<I_{dark,ij}>_{pix}}{\hoge<f_{s}(\lambda)>_{t}}\right\}\frac{{\delta}\beta_{sci}(\lambda,t)}{\hoge<\beta_{sci}(\lambda)>_{t}} .
\end{equation}
The residual error cannot be reduced through the calibration method and directly affects the transit measurements. Table \ref{tab:relative} shows the impacts of the decrease in the recapture rate on the relative spectro-photometric accuracies for the whole observing wavelength range of 3 to 22 \textmu m, the methane feature between 3.2 and 3.5 \textmu m, the ozone feature between 9 and 10 \textmu m, and the carbon dioxide feature between 13 and 17 \textmu m. The spectro-photometric accuracies for observations of the late- and middle-M type stars fully agree with the measurements of day-side emission and absorption features. This occurs because the timescales of the ramp for the late- and middle-M type stars are much longer than those for general-purpose spectrographs, owing to the relatively small number of photons falling on each pixel. However, it is difficult to perform the transmission spectroscopy and secondary eclipse of terrestrial planets orbiting early-M and late-K type stars even for the densified pupil spectrograph without any additional treatment.

\begin{deluxetable}{lccccc}[H]
\tablecaption{Values of $a$ and $\tau_{1}$ used for Equation (\ref{equ:beta}). \label{tab:beta}}
\tablehead{Items & Case 1\tablenotemark{c} & Case 2\tablenotemark{c} & Case 3\tablenotemark{c} & Case 4\tablenotemark{c} & HD 209458\tablenotemark{d}}
\startdata
$a$\tablenotemark{a} & 0.02 & 0.02 & 0.02 & 0.02 & 0.02 \\
$\tau_{1}$\tablenotemark{b} at 3 \textmu m (hour) & 135 & 37 & 4 & 1 & - \\
$\tau_{1}$\tablenotemark{b} at 10 \textmu m (hour) & 1185 & 379 & 51 & 18 & - \\
$\tau_{1}$\tablenotemark{b} at 15 \textmu m (hour) & 1854 & 605 & 83 & 29 & - \\
$\tau_{1}$\tablenotemark{b} at 22 \textmu m (hour) & 5307 & 1751 & 244 & 86 & - \\
$\tau_{1}$\tablenotemark{b} at 24 \textmu m (hour) & - & - & - & - & 10
\enddata
\tablenotetext{a}{$a$ was directly estimated from the result of \cite{2012ApJ...752...81C}.}
\tablenotetext{b}{$\tau_{1}$ was derived under an assumption that the timescale is determined only by the number of the photons falling on each pixel.}
\tablenotetext{c}{The cases of 1 to 4 correspond to the four simulation cases introduced in Section \ref{sec:simulation}, respectively.}
\tablenotetext{d}{HD 209458 was observed in the MIPS 24 \textmu m band \citep{2012ApJ...752...81C}.}
\end{deluxetable}

\begin{deluxetable}{lcccc}[H]
\tablecaption{Relative spectro-photometric errors arising from electron trapping for the four simulation cases. \label{tab:relative}}
\tablehead{Wavelength band & Case 1 (ppm) & Case 2 (ppm) & Case 3 (ppm) & Case 4 (ppm)}
\startdata
Whole observing wavelength range of 3 to 22 \textmu m & 0.9 & 9 & 1465 & 2798 \\
Methane band between 3.2 and 3.5 \textmu m & 0.2 & 2 & 315 & 210 \\
Ozone band between 9 and 10 \textmu m & 0.009 & 0.06 & 16 & 295 \\
Carbon dioxide band between 13 and 17 \textmu m & 0.009 & 0.07 & 19 & 276
\enddata
\end{deluxetable}

\section{Summary} \label{sec:summary}

We proposed a new method for mitigation and calibration of time-variation components present in the detector system, developing the densified pupil spectrograph that forms multiple spectra of divided primary mirrors on the detector plane to mitigate the impact of the wavefront error in the optical system on the transit spectroscopy. In addition, a number of the science pixels largely reduce the number of photons falling on each pixel and mitigate the ramp at the beginning of transit observation, the fallback after the ramp, and latent effect. The pixel-to-pixel time-variations can be also smoothed out, and the common time-variation components extracted, through average of a number of the science pixels. Here, focusing on the fact that the detector plane is optically conjugated to the primary mirror, a number of black-off pixels, called reference pixels, are formed by putting a mask on the pupil plane. Furthermore, placing an additional grating with a different angle from the one used for the science light on the focal plane, would result in part of the detector format being illuminated by only the background light (i.e., background pixels). Since the common time-variation components also affect the signals of the background and reference pixels, the components can be reconstructed from these pixels. The background light included in the science pixels can be removed through subtraction of background pixels from the science ones. Finally, the pixel-averaged science pixels are composed of the transit signal and random components. Note that the reduced random components after the calibration processes are enhanced by the shot noises of the background and reference pixels because the systematic components are reconstructed using the background and reference pixels.

We confirmed that the proposed method for calibration of the common systematic components over the entire detector plane is validated for the mid-infrared detector systems through mathematical analysis. We also performed numerical simulations to investigate whether the method effectively works under realistic assumptions. Sixty transit observations for transmission spectroscopy and secondary eclipse of terrestrial planets orbiting four host stars with the different effective temperatures of 2,500, 3,000, 3,500, and 4,000 K at 10 pc were performed with the densified pupil spectrograph mounted on the Origins Space Telescope having a 9.3 m primary mirror. We found that the method successfully reduced the systematic components into the random ones and nearly reached the photon-noise-limited performance for the temperatures of 2,500, 3,000 and 3,500 K. The proposed method contributes to the measurement of the biosignature and habitability of terrestrial planets orbiting M type stars. However, the systematic components for the late-K type stars remain because the second-order fluctuation affects the calibration method. In addition, the random components are slightly enhanced through the calibration process, compared to the shot noise associated to the raw data and limit the final spectro-photometric accuracy. Here, we succeeded in mitigation of the residual offset error through enhancement of the signal-to-noise ratios of the background and reference pixels.

Finally, we investigated impacts of the signal-to-noise ratios of the background and reference pixels. Higher ratios improve the calibration technique, except at the longer wavelengths of late- and middle-M type stars. We also studied the impact of time-variations of the background light and dark current upon the proposed method and found that the fluctuations of zodiacal light and dark current should be reduced down to 100 ppm. In terms of reducing the impact of the drift of the dark current during transit observation, we can measure the fluctuation of the detector temperature and calibrate the time-variation of the dark current instead of stabilizing the detector temperature. Note that the small fluctuation of the detector temperature can be measured using the relative-temperature-measurement mode of commercially available cryogenic sensors. Next, we investigated the ramp over the transit light curve due to the trapping effect and the estimation error of the offset voltage on the transit spectroscopy. Assuming that the timescale of the ramp is proportional to the number of the photons falling on each pixel, the timescale for the densified pupil spectrograph is expected to be much longer than those of general-purpose spectrographs are, because the densified pupil spectrograph forms multiple spectra on the detector plane. However, because bright stars with an effective temperature higher than 3,500 K at 10 pc generate a large number of photons falling on each pixel even for the densified pupil spectrograph, the relative spectro-photometric accuracies between wavelengths are limited by the ramp. The estimation error of the offset voltage also produces systematic components that influence the transit spectroscopy. By contrast, the effects of the estimation error of the offset voltage and the latent image upon the proposed method can be ignored unless the time-variation of the gain is not small.

We are building a proto-type of the densified pupil spectrograph optimized for the mid-infrared wavelength range and will demonstrate the proposed method, imitating an observing environment of the Origins Space Observatory.

\acknowledgments

We are sincerely grateful to Dr. Tom Greene and Dr. Robert McMurray for kindly teaching the Si:As IBC detector systems developed for infrared space telescopes. We also appreciate the WISE help desk for providing the WISE flat-field data. We acknowledge Tomoyasu Yamamuro for having discussions on the optical system for the background measurement. We also thank Dr. Hiroshi Shibai and Dr. Takahiro Sumi for having useful comments on the calibration method. Finally, we express our appreciation to the anonymous referee for numerous valuable comments on this study.

\appendix
\section{Parameters and assumptions} \label{sec:appendixA}
The parameters and their assumptions used for formulation and numerical simulations are compiled in Table A1.

\setcounter{table}{0}
\renewcommand{\thetable}{\Alph{table}}
\begin{longtable}[H]{|c|p{7cm}|p{7cm}|}
\caption{Parameter and their assumptions for formulation and numerical simulations.}
\label{tab:appendixA} \\ \hline
Parameters & Explanations & Assumptions \\ \hline \hline
\endfirsthead \hline
Parameters & Explanations & Assumptions \\ \hline \hline
\endhead
$\eta_{ij}$ & Quantum efficiency of (i, j) pixel & Fixed constant in each pixel \\ \hline
$\alpha_{ij}$ & Effective pixel gain of (i, j) pixel & Fixed constant in each pixel \\ \hline
$\beta_{ij}$ & Effective gain of (i, j) pixel related to electron trapping & Fixed constant in each pixel in Sections of \ref{sec:concept} to \ref{sec:simulation} and variable in Section \ref{sec:discussion} \\ \hline
$f_{s,ij}$ & Number of photons from an object source falling on (i, j) pixel & Common variable over the science pixels due to shot noise and planetary transit \\ \hline
$f_{z,ij}$ & Number of photons from zodiacal light falling on (i, j) pixel & Variable in each pixel due to shot noise \\ \hline
$t_{exp}$ & Exposure time & Fixed constant of 60 seconds \\ \hline
$C_{ij}$ & Capacitor capacitance in (i, j) pixel unit & Fixed constant in each pixel \\ \hline
$I_{dark,ij}$ & Dark current of (i, j) pixel & Variable in each pixel attaching shot noise \\ \hline
$I_{j}$ & Common current of j-th column & Variable in each column \\ \hline
$I_{k}$ & Common current in k-th gate & Variable in each gate \\ \hline
$A_{ij}$ & Transfer function of source follower in (i, j) pixel unit & Variable in each pixel \\ \hline
$A_{k}$ & Transfer function of k-th source follower & Variable in each gate \\ \hline
$V_{det,ij}$ & Detector bias voltage & Fixed constant in each pixel \\ \hline
$V_{dduc}$ & Drain voltage at source follower in each pixel unit & Fixed constant in each pixel \\ \hline
$V_{ddout}$ & Drain voltage at source follower in readout gate & Fixed constant in each gate \\ \hline
$V_{res,ij}$ & Reset voltage in (i, j) pixel unit & Variable in each pixel including shot noise present in (i, j) pixel \\ \hline
$v_{int,ij}$ & Integrated voltage of (i, j) pixel at source follower of (i, j) pixel unit & Variable in each pixel including shot noise present in (i, j) pixel \\ \hline
$v_{out,ijk}$ & Output voltage of (i, j) pixel at source follower at k-th readout gate & Variable in each pixel including shot noise present in (i, j) pixel \\ \hline
$v_{sci,ijk}$ & Output voltage of (i, j) science pixel with k-th readout gate & Variable in each pixel attaching \\ \hline
$v_{back,ijk}$ & Output voltage of (i, j) background pixel with k-th readout gate & Variable in each pixel attaching shot noise of zodiacal light \\ \hline
$v_{ref,ijk}$ & Output voltage of (i, j) reference pixel with k-th readout gate & Variable in each pixel attaching shot noise of dark current \\ \hline
$n_{sci}$ & Number of science pixels for spectrally-resolved band & Fixed constant of 120,000 \\ \hline
$n_{back}$ & Number of background pixels for spectrally-resolved band & Fixed constant of 120,000 \\ \hline
$n_{ref}$ & Number of reference pixels & Fixed constant of 760,000 \\ \hline
\end{longtable}

\end{document}